\documentclass[a4paper,11pt]{article}
\usepackage{a4wide,amsmath,amssymb,bbm}
\usepackage[english]{babel}
\usepackage[utf8]{inputenc}

\usepackage{color}

\usepackage{hyperref,enumerate}
\hypersetup{
    colorlinks,
    linkcolor={black},
    citecolor={black},
    urlcolor={black}
}

\usepackage{mathrsfs}

\newcommand{\be}{\begin{equation}}
\newcommand{\ee}{\end{equation}}

\newcommand{\dred}{\mathsf{D}}
\newcommand{\omegared}{\mathsf{w}}
\newcommand{\ered}{\mathsf{e}}

\makeatletter

\@addtoreset{equation}{section} \makeatother

\begin{document}

\hfill
\vspace{30pt}

\begin{center}
{\huge{\bf The first $\alpha'$-correction to homogeneous Yang-Baxter deformations using $O(d,d)$}}

\vspace{80pt}

Riccardo Borsato,$^a$ \ \ Alejandro Vilar L\'opez,$^a$ \ \  Linus Wulff$\, ^b$

\vspace{15pt}

{
\small {$^a$\it 
Instituto Galego de F\'isica de Altas Enerx\'ias (IGFAE), Universidade de  Santiago de Compostela, Spain}\\
\vspace{5pt}
\small {$^b$\it Department of Theoretical Physics and Astrophysics, Masaryk University, 611 37 Brno, Czech Republic}
\\
\vspace{12pt}
\texttt{riccardo.borsato@usc.es, alejandrovilar.lopez@usc.es, wulff@physics.muni.cz}}\\

\vspace{100pt}

{\bf Abstract}
\end{center}
\noindent
We use the $O(d,d)$-covariant formulation of supergravity familiar from Double Field Theory to find the first $\alpha'$-correction to (unimodular) homogeneous Yang-Baxter (YB) deformations of the bosonic string. A special case of this result gives the $\alpha'$-correction to TsT transformations. In a suitable scheme the correction comes entirely from an induced anomalous double Lorentz transformation, which is needed to make the two vielbeins obtained upon the YB deformation equal. This should hold more generally, in particular for abelian and non-abelian T-duality, as we discuss.

\pagebreak 
\tableofcontents

\setcounter{page}{1}


\section{Introduction}
Yang-Baxter deformations were originally constructed as deformations of the Principal Chiral Model and (super)coset sigma models with the interesting property that they preserve integrability \cite{Klimcik:2002zj,Klimcik:2008eq,Delduc:2013qra,Kawaguchi:2014qwa}. The deformations are built using an $R$-matrix which solves the classical Yang-Baxter equation (CYBE)
\begin{equation}
[RX,RY]-R([RX,Y]+[X,RY])=c^2[X,Y]\,,\qquad \forall X,Y\in\mathfrak g\,,
\label{eq:CYBE}
\end{equation}
where $c=0$ gives the standard CYBE equation and $c\neq0$ corresponds to the modified CYBE. We will consider only the $c=0$ case here, for which the deformed models are often called homogeneous YB models. It was shown in \cite{Hoare:2016wsk,Borsato:2016pas} that homogeneous deformations can be generated using non-abelian T-duality. One simply adds a closed, non-degenerate, $B$-field defined on a subalgebra of the isometry algebra and dualizes on that subalgebra.~\footnote{From this construction one obtains a deformation of non-abelian T-duality, but it is possible to show that a local field redefinition (i.e. a diffeomorphism in target space) plus a shift of the $B$-field permit to rewrite the result as a homogeneous YB deformation. See\cite{Hoare:2016wsk,Borsato:2016pas} for more details.} This construction means that these deformations can be defined for a general sigma model as long as it admits isometries that can be dualized. In particular the YB deformation of the Green-Schwarz superstring was constructed in \cite{Borsato:2018idb}. A special case of this deformation is when the isometries are abelian and in that case the deformed model is simply a T-duality - shift - T-duality (TsT) transformation \cite{Osten:2016dvf}, {which are usually called $\beta$-shifts or $\beta$-transformations in the context of $O(d,d)$.}

Just as in non-abelian T-duality \cite{Alvarez:1994np,Elitzur:1994ri}, these models may in principle have a Weyl anomaly. When the anomaly is present the target space fields do not solve the standard supergravity equations but a generalization of these \cite{Arutyunov:2015mqj,Wulff:2016tju}. Similar to the non-abelian T-duality case this anomaly is absent if one requires the $R$-matrix to satisfy a unimodularity condition \cite{Borsato:2016ose}. This is the case we consider here although unimodularity is not a necessary condition to avoid a Weyl anomaly \cite{Sakamoto:2018krs,Wulff:2018aku,Borsato:2018spz}.

The realization of homogeneous YB models using T-duality makes it natural to try to describe these models using the $O(d,d)$-covariant language of Double Field Theory (DFT), as was done starting with the work of \cite{Sakamoto:2017cpu}. In fact the YB deformations take the form of a so-called $\beta$-transformation \cite{Lust:2018jsx,Sakamoto:2018krs} in $O(d,d)$ language. This language is particularly useful since, as we will show in this paper, unimodular homogeneous YB deformations leave the generalized fluxes --- the basic building blocks in the $O(d,d)$-covariant formalism --- invariant (see also \cite{Lust:2018jsx,Bakhmatov:2018bvp}). With this observation it becomes very simple to prove that the deformed model solves the low-energy field equations, since those have an $O(d,d)$-covariant formulation in terms of the generalized fluxes. In fact the same is true for the first $\alpha'$-correction to these equations as shown in \cite{Baron:2017dvb}. Therefore it is also straightforward to argue that YB-deformed bosonic strings\footnote{Note that while here we consider only the bosonic string  for definiteness, very similar results hold for the heterotic string. In fact they can both be treated at the same time by introducing parameters that interpolate between the two as in for example \cite{Baron:2017dvb}. {Note that the relevant equations for the target space fields for the bosonic string are the type II supergravity equations with RR fields set to zero. Therefore we will often loosely refer to them  as the (super)gravity equations.}} are Weyl invariant at least up to two loops. The fact that all higher derivative corrections should respect the $O(d,d)$ structure suggest that this should even be true to all orders in $\alpha'$, although a complete proof that the string effective action can be written only in terms of generalized fluxes is not known to the authors. 

Naively this argument may seem to suggest that the YB-deformed backgrounds should not receive any $\alpha'$-corrections beyond those coming from the intrinsic $\alpha'$-dependence of the original (i.e. undeformed) background. But this is at odds with the results of \cite{Borsato:2019oip}, where non-trivial corrections were found working to second order in the expansion in the deformation parameter. When focusing on the class of TsT transformations, it is also at odds with the fact that abelian T-duality is known to receive $\alpha'$-corrections, as was shown in various works starting from~\cite{Tseytlin:1991wr,Panvel:1992he,Bergshoeff:1995cg,Haagensen:1997er,Kaloper:1997ux,Jack:1999av,Parsons:1999ze}, which would be expected to lead to corrections to TsT. As we will explain in more detail in the rest of the paper, the resolution is that while in the doubled formalism there is indeed no correction, corrections appear when one wants to go from the doubled formalism to a standard (super)gravity formulation. In order to do that one has to fix the double Lorentz gauge-invariance in such a way that the two vielbeins that naturally exist in the doubled formulation are set equal. This requires a certain double Lorentz transformation and --- given that the fields of the doubled formulation have an anomalous transformation\footnote{The fact that manifest $O(d,d)$ symmetry requires the fields to transform non-covariantly was clarified in the works \cite{Hohm:2011si,Hohm:2013jaa,Hohm:2014xsa,Hohm:2014eba}.} under double Lorentz transformations \cite{Marques:2015vua}  --- this induces an extra $\alpha'$-correction to the deformed background whose form we determine. A special case of our formula gives the $\alpha'$-correction to TsT transformations.

This discussion naturally connects also to the identification of $\alpha'$-corrections to abelian T-duality transformations, as mentioned above. We will discuss also this and comment on the comparison to the results of~\cite{Kaloper:1997ux}. Starting from the corrections to T-duality we will be able to provide an independent way to obtain $\alpha'$-corrections to TsT transformations, which does not make use of the double $O(d,d)$ formulation.

The outline of the paper is as follows. First we give a very brief introduction to the concepts needed from the $O(d,d)$-covariant formulation as used in DFT. In section \ref{sec:YB} we describe what Yang-Baxter deformations are in this language and show that they leave the generalized fluxes invariant. The $\alpha'$-correction to these deformations induced by the compensating anomalous Lorentz transformation is described in section \ref{sec:alpha-prime}. Section~\ref{sec:T-TsT} focuses on abelian T-duality and TsT transformations and we show that the results agree with those obtained using the $O(d,d)$-covariant formulation. We end with some concluding comments.

\section{\texorpdfstring{$O(d,d)$}{O(d,d)} covariant formulation of supergravity}
We will take inspiration from DFT and use the $O(d,d)$ covariant formulation of (super)gravity. In particular we will work with the so-called frame-like formulation of DFT \cite{Siegel:1993th,Siegel:1993xq,Hohm:2010xe} where the structure group is taken as two copies of the Lorentz group $O(1,d-1)\times O(d-1,1)$.  More details and references can be found in the reviews \cite{Aldazabal:2013sca,Hohm:2013bwa,Berman:2013eva}. However, unlike in DFT, we will always assume that the section condition is solved in the standard way $\partial_M=(0,\partial_m)$ so that we are really just working with a rewriting of supergravity. Here we will actually consider only the NSNS sector as appropriate for the bosonic string.

In the frame-like formulation one writes the generalized metric in terms of generalized (inverse) vielbeins
\begin{equation}
\mathcal H^{MN}=E_A{}^M\mathcal H^{AB}E_B{}^N\,,
\end{equation}
where $\mathcal H^{AB}$ is block diagonal with the usual Minkowski metric $\bar\eta=(-1,1,\ldots,1)$ in each block. Coordinate indices are raised and lowered with the $O(d,d)$ metric 
\begin{equation}
\eta^{MN}=\eta_{MN}=
\left(
\begin{array}{cc}
	0 & 1\\
	1 & 0
\end{array}
\right),
\end{equation}
and flat indices with the metric
\begin{equation}
\eta^{AB}=\eta_{AB}=
\left(
\begin{array}{cc}
	\bar\eta & 0\\
	0 & -\bar\eta
\end{array}
\right)\,.
\end{equation}
The generalized metric can be parameterized in the form
\begin{equation}
\mathcal H^{MN}=
\left(
\begin{array}{cc}
	G_{mn}-B_{mk}G^{kl}B_{ln} & B_{mk}G^{kn}\\
	-G^{mk}B_{kn} & G^{mn}
\end{array}
\right),
\label{eq:H}
\end{equation}
in terms of the usual metric $G$ and $B$-field. We take the generalized (inverse) vielbein to be
\begin{equation}
E_A{}^M=
\frac{1}{\sqrt2}
\left(
\begin{array}{cc}
e^{(+)a}{}_m-e^{(+)an}B_{nm} & e^{(+)am}\\
-e^{(-)}_{am}-e^{(-)}_a{}^nB_{nm} & e^{(-)}_a{}^m
\end{array}
\right)\,.
\label{eq:E}
\end{equation}
Here $e^{(\pm)}$ are two sets of vielbeins which transform independently  as $\Lambda^{(\pm)}e^{(\pm)}$ under the two Lorentz-group factors. To go to the standard supergravity picture one fixes a gauge $e^{(+)}=e^{(-)}=e$ leaving only one copy of the Lorentz-group.

An important object is the so-called generalized Weitzenb\"ock connection, defined in terms of the generalized vielbeins as
\begin{equation}
\Omega_{ABC}=E_A{}^M\partial_ME_B{}^NE_{CN}\,.
\label{eq:w-conn}
\end{equation}
From this the generalized fluxes are constructed as
\begin{equation}
\mathcal F_{ABC}=3\Omega_{[ABC]}\,,\qquad\mathcal F_A=\Omega^B{}_{BA}+2E_A{}^M\partial_M\hat d\,,
\end{equation}
where $\hat d$ is the generalized dilaton related to the standard one as $e^{-2\hat d}=e^{-2\Phi}\sqrt{-G}$. The importance of these objects comes from the fact that the generalized fluxes are scalars under generalized diffeomorphisms. This follows from the fact that a generalized diffeomorphism is implemented by the generalized Lie derivative which acts on a vector field as
\begin{equation}
\mathcal L_XY^M=X^N\partial_NY^M+(\partial^MX_N-\partial_NX^M)Y^N\,.
\end{equation}
The NSNS sector supergravity equations, or bosonic string low-energy effective equations, can be expressed in terms of the generalized fluxes only. To do this we first introduce the projectors
\begin{equation}
P_\pm=\frac12\left(\eta\pm\mathcal H\right)\,.
\end{equation}
Defining the following projections of the generalized fluxes 
\begin{equation}
\mathcal F^{(\pm)}_{ABC}=(P_{\mp})_A{}^D(P_{\pm})_B{}^E(P_{\pm})_C{}^F\mathcal F_{DEF}\,,\qquad\mathcal F^{(\pm)}_A=(P_\pm)_A{}^B\mathcal F_B\,,
\end{equation}
they take the form\footnote{Note that eq. (3.78) in \cite{Aldazabal:2013sca} is not correct, since for example the $P_+P_+$ projection does not vanish.}
\begin{align}
(P_+)_A{}^C(P_-)_B{}^D
\left[
\partial_C\mathcal F_D
-(\mathcal F^E-\partial^E)\mathcal F^{(-)}_{CDE}
+\frac14\mathcal F_C{}^{EF}\mathcal F_{DEF}
-\frac14(\mathcal F^2)_{CD}
\right]
=&\,0\,,
\\
\mathcal R=-4\partial_A\mathcal F^{(-)A}
+2\mathcal F_A\mathcal F^{(-)A}
+\frac14\mathcal F_A{}^{CD}\mathcal F_{BCD}\mathcal H^{AB}
-\frac{1}{12}\mathcal F^2
-\frac16\mathcal F_{ABC}\mathcal F^{ABC}=&\,0\,,
\end{align}
where $(\mathcal F^2)_{AB}=\mathcal F_{ACD}\mathcal H^{CE}\mathcal H^{DF}\mathcal F_{BEF}$ and $\mathcal F^2=\mathcal H^{AB}(\mathcal F^2)_{AB}$. The last line defines the generalized Ricci scalar and these equations of motion can be derived from the action
\begin{equation}
S=\int dX\,e^{-2\hat d}\mathcal R\,.
\end{equation}
Let us emphasize again that for us this is just a convenient rewriting of the usual bosonic string effective action and equations of motion at lowest order in $\alpha'$.

\section{Yang-Baxter deformations in \texorpdfstring{$O(d,d)$}{O(d,d)} language}\label{sec:YB}
We first need to show how to write YB deformations in $O(d,d)$ language at leading order in $\alpha'$, which will be needed later when discussing their $\alpha'$-corrections.
Under a YB deformation we have (e.g. \cite{Araujo:2017jkb,Borsato:2018idb})\footnote{We use a tilde to denote quantities after doing the deformation. We absorb the deformation parameter (usually denoted by $\eta$) into $\Theta$ to simplify the expressions.}
\begin{equation}\label{eq:SW}
G-B\rightarrow\tilde G-\tilde B=(G-B)(1+\Theta(G-B))^{-1}\,.
\end{equation}
The transformation of the dilaton is such that the generalized dilaton $\hat d$ is invariant. The transformation of $G$ and $B$ is equivalent to the following transformation of the generalized metric (\ref{eq:H})
\begin{equation}
\mathcal H\rightarrow\tilde{\mathcal H}=h^T\mathcal Hh\,,\qquad h_M{}^N=\delta_M{}^N+\Theta_M{}^N\,,
\end{equation}
where
\begin{equation}
\Theta_M{}^N=
\left(
\begin{array}{cc}
0 & \Theta^{mn}\\
0 & 0
\end{array}
\right)\,,\qquad\Theta^{mn}=k^m_rk^n_sR^{rs},
\end{equation}
where $k^m_r$ are Killing vectors of the undeformed background\footnote{{The assumption is that the Lie derivatives along $k^m_r$ of the metric, the $B$-field and the dilaton of the original background vanish. One could in principle relax the isometry condition on $B$ by demanding only that the Lie derivative of $H$ vanishes, but we will not consider this generalization here.}} and $R^{rs}$ is a constant anti-symmetric matrix satisfying (\ref{eq:CYBE}) with $c=0$ ($r,s$ are Lie algebra indices). Later we will show that if we just impose that $R$ is  constant and anti-symmetric,  the additional property of satisfying the CYBE~(\ref{eq:CYBE}) will have a natural interpretation. The generalized vielbein (\ref{eq:E}) then transforms as
\begin{equation}
E_A{}^M\rightarrow\tilde E_A{}^M=E_A{}^Nh_N{}^M\,.
\end{equation}
Note that the two sets of vielbeins in (\ref{eq:E}) transform differently, namely
\begin{equation}
\tilde e_a^{(\pm)m}=e_a^{(\pm)n}\left[\delta_n^m-(B_{nk}\mp G_{nk})\Theta^{km}\right]\,.
\label{eq:etilde}
\end{equation}
This means that if we start from an undeformed background in a gauge such that $e^{(+)}=e^{(-)}=e$, we will need to accompany the YB deformation by a generalized double Lorentz transformation. We will keep $\tilde e^{(+)}$ invariant and transform $\tilde e^{(-)}$ by
\begin{equation}
(\Lambda^{(-)})_a{}^b=\tilde\Lambda_a{}^b=[1+(G-B)\Theta]_a{}^c([1-(B+G)\Theta]^{-1})_c{}^b\,,
\label{eq:Lambda-}
\end{equation}
in order to preserve the gauge $\tilde e^{(+)}=\tilde e^{(-)}$. At the (super)gravity level this is of no concern since all objects transform covariantly, but when one considers $\alpha'$-corrections this transformation becomes important due to anomalous transformations of the fields, as we will discuss in the next section.

Note that this is a reformulation of YB deformations in the form of an $O(d,d)$ transformation, in fact $h$ has the form of a so-called $\beta$-transformation or $\beta$-shift. It is not a standard $O(d,d)$ transformation, such as the ones under which the DFT action is invariant, though. This is first of all because $\Theta^{mn}$ is (in general) not constant and second, and more importantly, because $\Theta^{mn}$ depends on the background itself since it is constructed using Killing vectors.  This is therefore \emph{not} a symmetry but a map of a background to another background, which is in fact a deformation of the first if we take $\Theta$ to be multiplied by a small parameter.

It follows from the transformation of the generalized vielbein that the generalized Weitzenb\"ock connection (\ref{eq:w-conn}) transforms as
\begin{align}
\tilde\Omega_{ABC}=&
E_A{}^Lh_L{}^M\partial_ME_B{}^NE_{CN}
+E_A{}^Lh_L{}^M\partial_Mh_K{}^N(h^{-1})_N{}^PE_B{}^KE_{CP}
\nonumber\\
=&
\Omega_{ABC}
+E_A{}^L\Theta_L{}^M\partial_ME_B{}^NE_{CN}
+E_A{}^Lh_L{}^M\partial_M\Theta_{KN}E_B{}^KE_C{}^N\,,
\end{align}
where we used the fact that any expression with two $\Theta$'s contracted (with $\eta_{MN}$) vanishes. Now we use the fact that $k_r$ generate isometries, i.e. the generalized Lie derivative of $E_A{}^M$ and $\hat d$ along $k_r$ vanish\footnote{Recall that $\hat d$ is a density rather than a scalar, hence the non-zero RHS in the second equation. {Note also that we are assuming the vielbeins and not just the metric to be invariant. This assumption was also made in \cite{Borsato:2018idb}, whose derivation we rely on, but it should be possible to relax it. We comment more on this in the next section.} }
\begin{equation}
k_r^L\partial_LE_A{}^M+(\partial^Mk_{rL}-\partial_Lk_r^M)E_A{}^L=0\,,\qquad k_r^K\partial_K\hat d=\frac12\partial_Kk_r^K\,.
\label{eq:gen-iso}
\end{equation}
Using this fact one finds that the change of the generalized flux $\mathcal F_{ABC}$ is proportional to the YB equation for $R$ in the form
\begin{equation}
\Theta^{M[K}\partial_M\Theta^{LN]}=0\,.
\end{equation}
Therefore $\mathcal F_{ABC}$ is invariant under a YB deformation (see also \cite{Lust:2018jsx,Bakhmatov:2018bvp}\footnote{{This is however at odds with \cite{Catal-Ozer:2019tmm}.}}). For $\mathcal F_A$ we find
\begin{equation}
\tilde{\mathcal F}_A=
{\mathcal F}_A
-\Theta_L{}^K\partial_KE_A{}^L
-\partial_K\Theta_L{}^KE_A{}^L
+2E_A{}^N\Theta_N{}^M\partial_M\hat d\,,
\end{equation}
and using (\ref{eq:gen-iso}) we find
\begin{equation}
\tilde{\mathcal F}_A=\mathcal F_A+E_A{}^M\Delta\mathcal F_M\,,\qquad
\Delta\mathcal F_M
=
\left(
\begin{array}{c}
-2\nabla_n\Theta^{mn}
\\
0
\end{array}
\right)\,.
\end{equation}
Therefore $\mathcal F_A$ is invariant precisely when the $R$-matrix is unimodular, since $\nabla_n\Theta^{mn}\propto f^t_{rs}R^{rs}$, and $f^t_{rs}R^{rs}=0$ is the unimodularity condition of~\cite{Borsato:2016ose}.

We have therefore shown that the generalized fluxes are invariant under unimodular YB deformations. In fact their derivatives are also invariant since for example
\begin{equation}
\partial_A\mathcal F_B=E_A{}^M\partial_M\mathcal F_B\rightarrow\partial_A\mathcal F_B-E_A{}^N\Theta_N{}^M\partial_M\mathcal F_B=\partial_A\mathcal F_B\,,
\end{equation}
because $k_r^M\partial_M\mathcal F_B=\mathcal L_{k_r}\mathcal F_B=0$ by isometry.

Since the (NSNS sector) supergravity equations of motion can be cast in terms of the generalized fluxes and their derivatives, this is enough to conclude that they are invariant under unimodular YB deformations. In other words such YB deformations map SUGRA solutions to SUGRA solutions. Moreover, also the first $\alpha'$-correction to the bosonic string equations can be cast in terms of the generalized fluxes and their derivatives, and therefore our argument shows that in fact the YB deformation preserves  Weyl invariance at least to two loops.\footnote{The action was written in terms of the fluxes in \cite{Baron:2017dvb}. But the variation of the generalized fluxes are again expressed in terms of the generalized fluxes which shows that the equations of motion are also expressed in this way, which is all we need.} In fact one would expect that all $\alpha'$-corrections to the equations can be expressed in $O(d,d)$ covariant form, which probably means they can be written only in terms of the generalized fluxes and their derivatives. If this is the case then our argument implies that YB deformations of the bosonic string preserve Weyl-invariance to all loops, i.e. they map a consistent bosonic string to another consistent bosonic string to all orders in $\alpha'$.

\section{The \texorpdfstring{$\alpha'$}{alpha'}-correction to YB deformations}\label{sec:alpha-prime}
Our general argument above has shown that YB deformations preserve two-loop Weyl invariance for the bosonic string. In fact they seem to require no additional $\alpha'$-corrections to the background besides those that are induced from the corrections to the original background. Here we want to understand how this fits with the results of \cite{Borsato:2019oip} where additional $\alpha'$-corrections were found for YB deformations. The resolution is that the additional $\alpha'$-corrections are indeed absent in the $O(d,d)$ covariant approach, but when one goes down to a standard supergravity formulation one has to fix the double Lorentz symmetry by  fixing $e^{(+)}=e^{(-)}=e$. The double Lorentz transformation required to do this induces, via the \emph{anomalous} transformation of the generalized vielbein at order $\alpha'$, additional $\alpha'$-corrections to the YB deformed model. Let us now see how this works.

It was shown in \cite{Marques:2015vua} that at order $\alpha'$ the generalized vielbein acquires an anomalous transformation under (double) Lorentz transformations. The transformation of the vielbein is given by\footnote{We are specifying here to the case of the bosonic string by setting $a=b=-\alpha'$ in the formulas of \cite{Marques:2015vua}.}
\begin{equation}
\delta E_A{}^M=-\lambda_A{}^BE_B{}^M+\alpha'\hat\delta_\lambda E_A{}^M\,,\qquad
\hat\delta_\lambda E_A{}^M=\left(\partial^{(-)}_{[A}\lambda_C{}^D\mathcal F^{(-)}_{B]D}{}^C-\partial^{(+)}_{[A}\lambda_C{}^D\mathcal F^{(+)}_{B]D}{}^C\right)E^{BM}\,,
\end{equation}
where $\lambda_C{}^D$ are parameters of an infinitesimal double Lorentz transformation and the second term is the anomalous piece. Note that we have defined the projected derivatives $\partial^{(\pm)}_A=(P_\pm)_A{}^B\partial_B$. After fixing the gauge $e^{(+)}=e^{(-)}=e$ the non-zero components of $\mathcal F^{(\pm)}$ are \cite{Marques:2015vua}
\begin{equation}
\mathcal F^{(+)}_M{}^{ab}=
\frac12\left(
\begin{array}{c}
G^{mn}\omega^{(+)ab}_n\\
-(1-BG)_m{}^n\omega^{(+)ab}_n
\end{array}
\right)\,,
\qquad
\mathcal F^{(-)}_{Mab}=
\frac12\left(
\begin{array}{c}
G^{mn}\omega^{(-)}_{nab}\\
(1+BG)_m{}^n\omega^{(-)}_{nab}
\end{array}
\right)\,,
\end{equation}
where $\omega_m^{(\pm)cd}=\omega_m{}^{cd}\pm\frac12H_m{}^{cd}$, {and the spin-connection is related to the vielbein and the Christoffel symbols $\Gamma_{mn}^p$ as 
\be
\omega_{mc}{}^d=e_c{}^n\partial_me_n{}^d-\Gamma_{mn}^pe_c{}^ne_p{}^d.
\ee
}
This leads to the anomalous infinitesimal transformations\footnote{The bar on the fields is to emphasize that these are the fields coming from the doubled formulation and which have an anomalous Lorentz transformation. Below we will define unbarred fields that transform covariantly.}
\begin{align}
\hat\delta\bar G_{mn}=&-\frac12\partial_{(m}\lambda^{(+)cd}\omega^{(+)}_{n)cd}-\frac12\partial_{(m}\lambda^{(-)cd}\omega^{(-)}_{n)cd}\,,\\
\hat\delta\bar B_{mn}=&\frac12\partial_{[m}\lambda^{(+)cd}\omega^{(+)}_{n]cd}-\frac12\partial_{[m}\lambda^{(-)cd}\omega^{(-)}_{n]cd}\,.
\label{eq:anom-trans}
\end{align}
Of course, after fixing the gauge $e^{(+)}=e^{(-)}=e$ only the transformations with $\lambda^{(+)}=\lambda^{(-)}=\lambda$ remain and the anomalous Lorentz transformations of the fields become
\begin{align}
\hat\delta\bar G_{mn}=&-\partial_{(m}\lambda^{cd}\omega_{n)cd}\,,\\
\hat\delta\bar B_{mn}=&\frac12\partial_{[m}\lambda^{cd}H_{n]cd}\,.
\end{align}
We see from these expressions that we can define new fields that transform non-anomalously by\footnote{We have included an extra shift of $G_{mn}$ by $H^2_{mn}$ to go to the scheme of Metsaev and Tseytlin (MT), see \cite{Marques:2015vua} and appendix~\ref{sec:schemes}.}
\begin{align}
G_{mn}^{(\mathrm{MT})}=&\bar G_{mn}+\alpha'\left(\frac12\omega_{mcd}\omega_n{}^{cd}+\frac38H_{mkl}H_n{}^{kl}\right)\,,\\
B_{mn}^{(\mathrm{MT})}=&\bar B_{mn}+\frac{\alpha'}{2}H_{cd[m}\omega_{n]}{}^{cd}\,.
\end{align}
The explicit non-covariant terms are constructed such as to cancel the anomalous Lorentz transformations. Notice that the above redefinitions also fix the \emph{finite} form of the anomalous Lorentz transformations of $\bar G,\bar B$.


\subsection{Compensating anomalous transformation}
When we are dealing with the YB deformation it is crucial to remember the compensating double Lorentz transformation needed to make $\tilde e^{(+)}=\tilde e^{(-)}$ given by (\ref{eq:Lambda-}). Setting $\lambda^{(+)}=0$ and $\lambda^{(-)}=\tilde\lambda$ in (\ref{eq:anom-trans}) we find that this induces an extra transformation of the fields at order $\alpha'$ given by\footnote{{In (\ref{eq:anom-trans}) we assumed  $e^{(+)}=e^{(-)}=e$ and we were doing a double Lorentz transformation from that starting point. Here we can use the same logic, assuming that we start from the gauge  $\tilde e^{(+)}=\tilde e^{(-)}=\tilde e$ for a YB deformation and go back to the situation where  $\tilde e^{(+)}=\tilde e$ and $\tilde e^{(-)}=\tilde\Lambda^T\tilde e$ as in~\eqref{eq:etilde}.} In this way we construct the inverse of the anomalous transformation we want. We remind that $\omega_m^{(\pm)cd}=\omega_m{}^{cd}\pm\frac12H_m{}^{cd}$, so that the $(\pm)$ on the torsionful spin-connection should not be confused with the $(\pm)$ on the two vielbeins coming from DFT. Setting $\lambda^{(+)}=0$ means that for the deformed model we take $\tilde e=\tilde e^{(+)}$.}
\begin{align}
\hat\delta\bar G_{mn}=-\frac12\partial_{(m}\tilde\lambda^{cd}\tilde\omega^{(-)}_{n)cd}\,,\qquad
\hat\delta\bar B_{mn}=-\frac12\partial_{[m}\tilde\lambda^{cd}\tilde\omega^{(-)}_{n]cd}\,.
\end{align}
We now need the finite form of the transformation since we are doing a finite transformation $\tilde\Lambda=e^{\tilde\lambda}$ given by (\ref{eq:Lambda-}). To find it we use the same strategy as above. We redefine $G$ and $B$ by terms involving the spin connection in such a way that the new fields do not have any anomalous transformation. From this one can then read off the finite form of the transformation.

For $\bar G_{mn}$ this is easily done by noting that $\bar G_{mn}+\frac{\alpha'}{4}\tilde\omega_m^{(-)cd}\tilde\omega^{(-)}_{ncd}$ is invariant under the above transformation and so the finite transformation for $\bar G$ is\footnote{Recall that we are computing minus the anomalous transformation we are after.}
\begin{equation}
\delta_{\mathrm{comp}}\bar G_{mn}=
-\frac12[\tilde\Lambda\partial_{(m}\tilde\Lambda^T]^{cd}\tilde\omega^{(-)}_{n)cd}
+\frac14[\tilde\Lambda\partial_m\tilde\Lambda^T]^{cd}[\tilde\Lambda\partial_n\tilde\Lambda^T]_{cd}\,.
\end{equation}
For $\bar B_{mn}$ things are more subtle because a similar term $\frac14\tilde\omega_{[m}^{(-)cd}\tilde\omega^{(-)}_{n]cd}$ vanishes by anti-symmetry. The part involving $H$ in $\omega^{(-)}$ can be integrated as before, while the part involving $\omega$ can be found by the following trick. Consider the anomalous transformation of $H=dB$ instead. One finds that $H$ transforms like the Chern-Simons form for $\omega$
\begin{equation}
\hat\delta\bar H=-\frac14\delta\mathrm{CS}(\omega)=-\frac14\delta\mathrm{tr}(\omega d\omega+\frac23\omega\omega\omega)\,.
\end{equation}
The finite transformation of the CS form is
\begin{equation}
\delta\mathrm{CS}(\omega)=d(\tilde\Lambda d\tilde\Lambda^T\omega)-\frac13\mathrm{tr}(\tilde\Lambda^Td\tilde\Lambda\tilde\Lambda^Td\tilde\Lambda\Lambda^Td\tilde\Lambda)\,.
\end{equation}
This implies that the transformation of $B$ can be taken to be
\begin{equation}
\delta_{\mathrm{comp}}\bar B_{mn}=
-\frac12[\tilde\Lambda\partial_{[m}\tilde\Lambda^T]^{cd}\tilde\omega^{(-)}_{n]cd}
+B^{\mathrm{WZW}}_{mn}\,,
\end{equation}
where $B^{\mathrm{WZW}}$ is defined by
\begin{equation}
dB^{\mathrm{WZW}}=-\frac{1}{12}\mathrm{tr}(\tilde\Lambda^Td\tilde\Lambda\tilde\Lambda^Td\tilde\Lambda\tilde\Lambda^Td\tilde\Lambda)\,.
\label{eq:HWZW}
\end{equation}
Now that we have found the pieces induced by the compensating double Lorentz transformation we are ready to write the $\alpha'$-correction to the YB-transformed metric and $B$-field.

\subsection{The correction to Yang-Baxter deformations}
Putting everything together the $\alpha'$-correction to the YB-deformed background in the scheme of Hull and Townsend is\footnote{See appendix~\ref{sec:schemes} for the field redefinitions connecting all schemes. Here we set the parameter $q$ of Hull and Townsend to zero.}
\begin{align}
\delta(\tilde G-\tilde B)_{mn}^{(\mathrm{HT})}
=&
\frac12\tilde\omega^{(-)}_{mcd}\left(\tilde\omega^{(+)}_n{}^{cd}-[\tilde\Lambda\partial_n\tilde\Lambda^T]^{cd}\right)
+\frac14\partial_m\tilde\Lambda^{cd}\partial_n\tilde\Lambda_{cd}
-B^{\mathrm{WZW}}_{mn}
+\delta'(\tilde G-\tilde B)_{mn}\,,
\label{eq:non-cov}
\\
\delta\tilde\Phi^{(\mathrm{HT})}=&
\frac14\tilde G^{kl}\delta\tilde G_{kl}
+\frac{1}{48}(\tilde H^2-H^2)\,.\label{eq:dil}
\end{align}
{The correction to the dilaton follows from the fact that in the HT scheme when $\Phi'=\Phi^{(\mathrm{HT})}-\frac{1}{48}\alpha'H^2$ the combination $e^{-2\Phi'}\sqrt{-G}$ is invariant under YB deformations, up to order $\alpha'$ included\footnote{Notice that $\Phi'$ is in fact the dilaton in the HT scheme at $q=1/6$.}~\cite{Borsato:2019oip}.} The term $\delta'(\tilde G-\tilde B)$ takes into account the scheme-change of the undeformed background\footnote{{For the same reason we have also a $\frac12\tilde\omega^{(-)}_m{}^{cd}\tilde\omega^{(+)}_{ncd}$ term in the correction above, generated by the scheme-change after the deformation.}}
\begin{equation}
\delta(G-B)_{mn}=-\frac12\omega^{(-)}_m{}^{cd}\omega^{(+)}_{ncd}\,,
\end{equation}
needed to relate the HT scheme to the $O(d,d)$ covariant scheme (see appendix \ref{sec:schemes}) and it takes the form
\begin{equation}
\delta'(\tilde G-\tilde B)_{mn}=[(1+(G-B)\Theta)^{-1}\delta(G-B)(1+\Theta(G-B))^{-1}]_{mn}\,.
\end{equation}
Note that in addition to this, one has the $\alpha'$-corrections to the original background, which will need to be included in~\eqref{eq:SW} and will therefore induce a term of the same form --- where now $\delta(G-B)$ is the correction to the original background.

It is important to stress that our derivation assumes that the $B$-field and vielbein of the undeformed background are invariant under the isometries generated by the Killing vectors entering $\Theta$. When there is no gauge where this is possible, equations~\eqref{eq:non-cov} and~\eqref{eq:dil} do not necessarily lead to a background solving the $\alpha'$-corrected supergravity equations. See however the next subsection.

The spin connection for the YB deformed background entering these expressions is computed using the vielbein $\tilde e=\tilde e^{(+)}$ defined in (\ref{eq:etilde}) and is given by
\begin{equation}
\tilde\omega_m{}^{ab}(\tilde e^{(\pm)})
=
\omega_m{}^{ab}
+\nabla_m[(B\mp G)\Theta]^{[a|k|}([1-(B\mp G)\Theta]^{-1})_k{}^{b]}
-\tilde e^{(\pm)[a|k|}\tilde e^{(\pm)b]l}\nabla_k\tilde G_{lm}\,.
%
\end{equation}

To see that (\ref{eq:non-cov}) and (\ref{eq:dil}) reproduces the results found in \cite{Borsato:2019oip} one sets $B=0$ and expands to order $\Theta^2$ obtaining
\begin{align}
\delta\tilde G_{mn}=&
-\nabla_m\Theta^{cd}\nabla_c\Theta_{dn}
-\nabla_n\Theta^{cd}\nabla_c\Theta_{dm}
+\mathcal O(\Theta^4)
\\
\delta\tilde B_{mn}=&2\partial_{[m}\left(\omega_{n]}{}^{cd}\Theta_{cd}\right)-\Theta_{cd}R_{mn}{}^{cd}
+\mathcal O(\Theta^3)
\\
\delta\tilde\Phi=&
\frac{1}{16}\nabla^m\Theta_{cd}\nabla_m\Theta_{cd}
-\frac{3}{8}\nabla^m\Theta^{cd}\nabla_c\Theta_{dm}
+\mathcal O(\Theta^4)\,,
\end{align}
which, up to a diffeomorphism and $B$-field gauge transformation, is the same as in \cite{Borsato:2019oip}. Note that one has to use the fact that the isometry of the vielbein implies that
\begin{equation}
i_k\omega^{ab}=-\nabla^ak^b\,.
\label{eq:ikomega}
\end{equation}

It is worth noting that in the case of a single TsT transformation the correction simplifies. Recall that, given two isometric coordinates $y_1,y_2$, a TsT transformation is implemented by the sequence of T-duality $y_1\to T(y_1)$ followed by a shift $y_2\to y_2-\eta T(y_1)$ and by another T-duality $T(y_1)\to y_1$. It is understood as a special case of YB with $\Theta=\eta k_1\wedge k_2$, where $k_i=\partial_{y_i}$ are Killing vectors. The above correction simplifies in the TsT case since $B^{\mathrm{WZW}}$ vanishes. This follows by noting that $\tilde\Lambda=1+2\Theta([1-(B+G)\Theta]^{-1})$ which means that when $\Theta$ has rank 2 the Lorentz transformation is only non-trivial in a $2\times2$ block. In this block it is $e^\lambda$ with $\lambda$ an anti-symmetric $2\times2$ matrix. Since such a matrix only has one independent component, the RHS of (\ref{eq:HWZW}) vanishes.

\subsection{Manifestly covariant form of the correction}
The expression (\ref{eq:non-cov}) for the $\alpha'$-correction is not manifestly covariant but one can show that it is nevertheless covariant. We start by noting that\footnote{{Here and in the following the covariant derivative is the one for the undeformed metric $G$. Moreover, unless written explicitly otherwise, one should use the undeformed vielbein to go from curved to flat indices.}}
\begin{equation}
\tilde\omega_m'^{(\pm)ab}
=
-\tilde e^{[a|k|}\tilde e^{b]l}\nabla_k(\tilde G\pm\tilde B)_{ml}
+\frac12\tilde e^{[a|k|}\tilde e^{b]l}\nabla_m(\tilde G\pm\tilde B)_{kl}
-\nabla_m[(G-B)\Theta]^{[a|k|}([1+(G-B)\Theta]^{-1})_k{}^{b]}
\end{equation}
where $\tilde\omega'^{(\pm)}=\tilde\omega^{(\pm)}-\omega$. With a bit of algebra one finds{
\begin{align}
\tilde\omega'^{(+)}_{mcd}
=&
\frac12\nabla_mB_{cd}
-\frac12[(G-B)\nabla_m\Theta(G+B)]_{cd}
+[1+(G-B)\Theta]_{[c|k|}(\nabla^kB(1+\Theta(G-B))^{-1})_{d]m}
\nonumber\\
&{}
+[1+(G-B)\Theta]_{[c|k|}((G-B)\nabla^k\Theta(1+(G-B)\Theta)^{-1}(G-B))_{d]m}
\nonumber\\
=&
\frac12H_{mcd}
-X^{(+)}_{kcd}[(G-B)(1+\Theta(G-B))^{-1}]^k{}_m\,,
\end{align}
where we have defined\footnote{When the vielbeins are invariant under the isometries, (\ref{eq:ikomega}) gives $\omega^{(\pm)}_{lcd}\Theta^l{}_k=X^{(\pm)}_{kcd}$.}
\begin{equation}
X_{kcd}^{(\pm)}=\frac12\nabla_k\Theta_{cd}-\nabla_{[c}\Theta_{d]k}\pm\frac12H_{cdl}\Theta^l{}_k
\end{equation}
and we used the YB equation in the last term of the first expression and also the isometry of $B$ in the next to last term. A similar calculation gives
\begin{equation}
[\tilde\Lambda^T\tilde\omega'^{(-)}_m\tilde\Lambda+\tilde\Lambda^T\nabla_m\tilde\Lambda]_{cd}
=
-\frac12H_{mcd}
+X^{(-)}_{kcd}[(G+B)(1-\Theta(G+B))^{-1}]^k{}_m\,.
\end{equation}
Using these expressions we find that (\ref{eq:non-cov}) can be written instead as
\begin{align}
\label{eq:cov}
\delta(\tilde G-\tilde B)_{mn}=&
-\tfrac14\nabla_m\tilde\Lambda^{cd}\nabla_n\tilde\Lambda_{cd}
-B^{\mathrm{cov-WZW}}_{mn}
+\tfrac12[\nabla_m\tilde\Lambda\tilde\Lambda^T]^{cd}\left(X^{(+)}_{kcd}(\tilde G-\tilde B)^k{}_n-\tfrac12H_{ncd}\right)
\nonumber\\
&{}
+\tfrac12[\tilde\Lambda^T\nabla_n\tilde\Lambda]^{cd}\left(X^{(-)}_{kcd}(\tilde G-\tilde B)_m{}^k-\tfrac12H_{mcd}\right)
\\
&{}
+\tfrac14\left(\tilde\Lambda^c{}_e\tilde\Lambda^d{}_f-\delta^c_e\delta^d_f\right)
\Big[
(\tilde G-\tilde B)_m{}^kX^{(-)ef}_kH_{ncd}
+(\tilde G-\tilde B)^k{}_nX^{(+)}_{kcd}H_m{}^{ef}
\nonumber\\
&{}
\qquad\qquad\qquad\qquad-2(\tilde G-\tilde B)_m{}^kX^{(-)ef}_kX^{(+)}_{lcd}(\tilde G-\tilde B)^l{}_n
-\tfrac12H_m{}^{ef}H_{ncd}
\Big]\,,
\nonumber
\end{align}
where $\tilde G-\tilde B$ is given by (\ref{eq:SW}) and we have defined
\begin{equation}
B^{\mathrm{cov-WZW}}_{mn}
=
B^{\mathrm{WZW}}_{mn}
-\frac12\mathrm{tr}\left(\omega_{[m}\tilde\Lambda\nabla_{n]}\tilde\Lambda^T\right)
+\frac12\mathrm{tr}\left(\omega_{[m}\tilde\Lambda^T\partial_{n]}\tilde\Lambda\right)\,.
\end{equation}
The correction to the dilaton is still given by (\ref{eq:dil}). All terms except $B^{\mathrm{cov-WZW}}$ are now manifestly covariant.} For the latter the identity
\begin{align}
\mathrm{tr}\left([\tilde\Lambda^T\nabla\tilde\Lambda]^3\right)
-\mathrm{tr}\left([\tilde\Lambda^Td\tilde\Lambda]^3\right)
=&
-\frac32d\mathrm{tr}\left(\omega[d\tilde\Lambda\tilde\Lambda^T+\tilde\Lambda^Td\tilde\Lambda]\right)
-\frac32\nabla\mathrm{tr}\left(\omega[\nabla\tilde\Lambda\tilde\Lambda^T+\tilde\Lambda^T\nabla\tilde\Lambda]\right)
\nonumber\\
&{}
+3\mathrm{tr}\left(R[\nabla\tilde\Lambda\tilde\Lambda^T+\tilde\Lambda^T\nabla\tilde\Lambda]\right)\,,
\end{align}
where {$R=d\omega+\omega\wedge \omega$} is the curvature 2-form, implies
\begin{align}
dB^{\mathrm{cov-WZW}}
=
-\frac{1}{12}\mathrm{tr}\left([\tilde\Lambda^T\nabla\tilde\Lambda]^3\right)
+\frac14\mathrm{tr}\left(R[\nabla\tilde\Lambda\tilde\Lambda^T+\tilde\Lambda^T\nabla\tilde\Lambda]\right)\,.
\end{align}
Therefore also the transformation of $B$ is covariant (up to $B$-field gauge transformations).

{The manifestly covariant form of the correction given by (\ref{eq:cov}) is actually more useful than the original form (\ref{eq:non-cov}). The reason is that  our derivation has assumed that the vielbeins are invariant under the isometries used to construct $\Theta$, and therefore (\ref{eq:non-cov}) is valid only in this case. Being covariant, (\ref{eq:cov}) is valid also when the vielbeins are not invariant under the isometries, as long as there exists a gauge in which they are invariant. In fact, even though it is not guaranteed by our construction, these expressions can be valid more generally, i.e. even in cases where it is not possible to find a gauge in which the vielbeins are invariant. We mention one such example below.
}

\subsection{Tests on examples}
{
We have tested the formulas~\eqref{eq:non-cov},~\eqref{eq:dil} for $\alpha'$-corrections to YB deformations on a number of examples, to check that they generate backgrounds solving the $\alpha'$-corrected supergravity equations.	First we worked out deformations of a Bianchi II background first considered in~\cite{Borsato:2019oip}. We tested our results both on the abelian deformations $\Theta=k_1\wedge k_4$ and $\Theta=k_2\wedge k_3$, and on the non-abelian deformation $\Theta=k_1\wedge k_4+k_2\wedge k_3$. We refer to~\cite{Borsato:2019oip} for the $\alpha'$-correction of the undeformed background and for the definition of the Killing vectors $k_i$, {whose non-trivial commutation relations are just $[k_1,k_2]=k_3$}. On this Bianchi II example we find that $B^{\mathrm{WZW}}$ is trivial even when considering the non-abelian deformation.

We worked out also deformations of the pure NSNS $AdS_3\times S^3$ background.\footnote{The $\alpha'$ corrections of the undeformed background are simply obtained by multiplying metric and $B$-field by $1+2\alpha'$ on the $AdS$ part and by $1-2\alpha'$ on the sphere part.} Its YB deformations were classified in~\cite{Borsato:2018spz}. We worked out various abelian deformations corresponding to TsT transformations on the sphere, on $AdS$, or mixing the two spaces. We worked out also the non-abelian deformation generated by $\Theta = (k_0+\bar k_0)\wedge k_s + k_+\wedge \bar k_-$. Here $k_s$ is a Killing vector on the sphere and we refer to~\cite{Borsato:2018spz} for the definitions we use for the $AdS$ Killing vectors. In this case we cannot immediately apply~\eqref{eq:non-cov} because it is not possible to find a vielbein for the $AdS_3$ metric that is invariant under all the isometries entering $\Theta$. We can anyway obtain $\alpha'$-corrections for this non-abelian deformation if we use the covariant formula~\eqref{eq:cov}. Alternatively, we can interpret this particular deformation as a non-commuting sequence of TsT transformations. Doing so, we can first work out the $\alpha'$-corrected abelian deformation generated by $\Theta =  k_+\wedge \bar k_-$, and after doing that we can work out the abelian deformation $\Theta = (k_0+\bar k_0)\wedge k_s$.\footnote{After doing the first abelian deformation, and before applying the second one, one has to carefully choose the vielbein such that it is invariant under the $k_0+\bar k_0$ isometry. At this stage it is not necessary anymore to impose the invariance under $k_+, \bar k_-$, which is what saves the day in this approach.}
}

\section{T-duality and TsT transformations}\label{sec:T-TsT}
Abelian T-duality transformations are another class of $O(d,d)$ transformations and we can follow exactly the reasoning in section~\ref{sec:alpha-prime} to obtain their $\alpha'$-corrections. When we remain in the non-covariant scheme that comes from  DFT, the corrections to the dualized metric and $B$-field will be given again by the formula  (a hat on the field is used to denote the T-dualization)
\begin{equation}
\delta(\widehat G-\widehat B)_{mn}
=
-\frac12\hat\omega^{(-)}_{mcd}(\hat\Lambda\partial_n\hat\Lambda^T)^{cd}
+\frac14\partial_m\hat\Lambda^{cd}\partial_n\hat\Lambda_{cd}
-B^{\mathrm{WZW}}_{mn}
\,,
\end{equation}
where now the Lorentz matrix is
\be
\hat\Lambda_a{}^b=\delta_a{}^b-2G_{yy}^{-1}e_{ya}e_y{}^b\,.
\ee
We are assuming that we are dualising along the coordinate $y$ and expressions for the corrections in other schemes will be obtained by implementing the relevant field redefinitions, see appendix~\ref{sec:schemes}.

In~\cite{Edelstein:2019wzg} $\alpha'$-corrections to the T-duality rules from the DFT formulation were also discussed. There however instead of writing the generic form of the corrections in terms of the finite form of the Lorentz transformation as above, it was noted that $\hat \Lambda$ reduces to a constant\footnote{It is $\hat \Lambda=\text{diag}(-1,1,\ldots,1)$ where the dualized coordinate is placed first.} when choosing a specific gauge for the vielbein\footnote{For curved indices we take $m=y,\mu$ and similarly we also have flat indices $a=\iota , \alpha$. We denote by $\ered_\mu{}^\alpha$ the vielbein for the reduced metric $g_{\mu\nu}$ appearing below.}
\be\label{eq:choice-E}
e_\mu{}^\alpha=\ered_\mu{}^\alpha\,,\qquad
e_y{}^\alpha=0\,,\qquad
e_\mu{}^\iota{}=e^\sigma V_\mu\,,\qquad
e_y{}^\iota=e^{\sigma}\,.
\ee
Here we are rewriting the fields in terms of fields of a dimensional reduction
\be
\begin{aligned}
ds^2 &= G_{mn}dx^mdx^n=g_{\mu\nu}dx^\mu dx^\nu +e^{2\sigma}(dy+V)^2,\\
B&= \frac{1}{2}B_{mn}dx^m\wedge dx^n=\frac{1}{2}b_{\mu\nu}dx^\mu\wedge dx^\nu+\frac{1}{2}W\wedge V + W \wedge dy\,,\\
\Phi&=\phi+\frac{1}{2}\sigma.
\end{aligned}
\ee
Since $\hat \Lambda$ is constant the anomalous Lorentz transformation is trivial in this gauge, and also the $\alpha'$-corrections to T-duality will be trivial.\footnote{Importantly, this statement is gauge dependent, in accordance with the fact that the scheme under discussion is not Lorentz-covariant. Covariant schemes such as HT or MT will not have this type of gauge ambiguity. } {(Note that while it is possible to avoid corrections for a single T-duality it is not possible in general for more than one T-duality, as shown in \cite{Eloy:2019hnl}.)}
In~\cite{Edelstein:2019wzg} this observation was used to obtain the $\alpha'$-corrections to the T-duality rules in the scheme of Bergshoeff and de Roo (BR)~\cite{Bergshoeff:1988nn,Bergshoeff:1989de}.
We use this result as a starting point to write below the T-duality rules to 2 loops in a family of different schemes.
\begingroup
\allowdisplaybreaks
\begin{align} \label{rulesGeneric}
\hat{\sigma} = & \, - \sigma + \left( a_1 - \frac{a_4}{2} + 2 a_5 + 2 \gamma_+ \right) (\dred \sigma)^2 - \frac{1}{8} \left(a_1 + 4 a_2 - a_5 - 2 \gamma_+ \right) \left( e^{2\sigma} V^{\lambda\rho} V_{\lambda\rho} + e^{-2\sigma} W^{\lambda\rho} W_{\lambda\rho}\right) - \nonumber \\
 & \, - \frac{1}{2} \left( \gamma_- - a_6 \right) V^{\lambda\rho} W_{\lambda\rho} ~ , \\
\hat{V}_\mu = & \,  W_\mu + \frac{1}{2} \left( \gamma_+ - b_3 + a_5 \right) W_\beta{}^\alpha \omegared_{\mu\alpha}{}^\beta + \frac{e^{2 \sigma}}{4} \left( - 4 a_2 + 2 b_1 + b_3 + \gamma_+ \right) h_{\mu\lambda\rho} V^{\lambda\rho} + \nonumber \\
 & \, + \frac{1}{4} \left(6 a_1 - a_4 + 4 a_5 + 4 b_1 - 2 b_2 + 4 b_3 + 4 \gamma_+ \right) W_{\mu p} \dred^p \sigma + \frac{1}{2} \left( a_4 - 2 b_2 \right) W_{\mu\rho} \dred^\rho \phi - \nonumber \\
 & \, - \frac{1}{2} \left( a_1 + 2 b_1 \right) \dred^\rho W_{\mu \rho} - \frac{1}{2} \left( \gamma_- - a_6 \right) \left( e^{2 \sigma} V_\beta{}^\alpha \omegared_{\mu\alpha }{}^\beta + \frac{1}{2} h_{\mu \lambda\rho} W^{\lambda\rho} - 2 e^{2 \sigma} V_{\mu \rho} \dred^\rho \sigma \right) ~ , \\
\hat{W}_\mu  = & \,  V_\mu  - \frac{1}{2} \left( \gamma_+ - b_3 + a_5 \right) V_\beta{}^\alpha \omegared_{\mu\alpha }{}^\beta - \frac{e^{-2 \sigma}}{4} \left( - 4 a_2 + 2 b_1 + b_3 + \gamma_+ \right) h_{\mu \lambda\rho} W^{\lambda\rho} + \nonumber \\
 & \, + \frac{1}{4} \left(6 a_1 - a_4 + 4 a_5 + 4 b_1 - 2 b_2 + 4 b_3 + 4 \gamma_+ \right) V_{\mu \rho} \dred^\rho \sigma - \frac{1}{2} \left( a_4 - 2 b_2 \right) V_{\mu \rho} \dred^\rho \phi + \nonumber \\
 & \, + \frac{1}{2} \left( a_1 + 2 b_1 \right) \dred^\rho V_{\mu \rho} + \frac{1}{2} \left( \gamma_- - a_6 \right) \left( e^{-2 \sigma} W_\beta{}^\alpha \omegared_{\mu\alpha }{}^\beta + \frac{1}{2} h_{\mu \lambda\rho} V^{\lambda\rho} + 2 e^{-2 \sigma} W_{\mu \rho} \dred^\rho \sigma \right) ~ , \\
\hat{\phi} = & \, \phi - \frac{1}{16} \left( a_1 - 4 a_2 - a_5 + 4 c_1 + 48 c_2 \right) \left( e^{2 \sigma} V_{\lambda\rho} V^{\lambda\rho} - e^{-2 \sigma} W_{\lambda\rho} W^{\lambda\rho} \right) + \nonumber \\
 & \, + \frac{1}{2} \left( a_1 - 8 c_1 + 2 c_4 \right) \dred^2 \sigma - \frac{1}{2} \left( a_4 - 4 c_3 - 4 c_4 \right) \dred_\rho \sigma \dred^\rho \phi ~ , \\
\hat{g}_{\mu\nu} = & \, g_{\mu\nu} - \frac{1}{2} \left( a_1 + 4 a_2 + a_5 \right) \left( e^{2 \sigma} V_{\mu\rho} V_\nu{}^\rho - e^{-2 \sigma} W_{\mu\rho} W_\nu{}^\rho \right) +  \nonumber \\
 & \, + \left(- 2 a_1 + a_4 \right) \dred_\mu \dred_\nu \sigma + 2 a_3 \dred_{(\mu} \sigma \dred_{\nu)} \phi ~ , \\
\hat{b}_{\mu\nu} = & \, b_{\mu\nu} - \frac{1}{2} \left( \gamma_+ - b_3 + a_5 \right) \left( V_\beta{}^\alpha \omegared_{[\mu\alpha}{}^\beta W_{\nu]} - W_\beta{}^\alpha \omegared_{[\mu\alpha}{}^\beta V_{\nu]} \right) + \frac{1}{2} \left( a_1 + 2 b_1 \right) \left( \dred^\rho W_{\rho[\mu} V_{\nu]} - \dred^\rho V_{\rho[\mu} W_{\nu]} \right) + \nonumber \\
 & \, + \frac{1}{4} \left( 4 a_2 - 2 b_1 - b_3 - \gamma_+ \right) \left( e^{2 \sigma} V_{[\mu} h_{\nu]\lambda\rho} V^{\lambda\rho} - e^{- 2 \sigma}  W_{[\mu} h_{\nu]\lambda\rho} W^{\lambda\rho} \right) + \left( 2 b_1 + b_2 \right) h_{\mu\nu \rho} \dred^\rho \sigma + \nonumber \\
 & \, + \frac{1}{4} \left( -6 a_1 + a_4 - 4 a_5 - 4 b_1 + 2 b_2 - 4 b_3 - 4 \gamma_+ \right) \left( V_{[\mu} W_{\nu]\rho} \dred^\rho \sigma + W_{[\mu} V_{\nu]\rho} \dred^\rho \sigma \right) - \nonumber \\
 & \, - \frac{1}{2} \left( a_4 - 2 b_2 \right) \left( V_{[\mu} W_{\nu]\rho} \dred^\rho \phi - W_{[\mu} V_{\nu]\rho} \dred^\rho \phi \right) - 2 b_3 V^\rho{}_{[\mu} W_{\nu]\rho} + \nonumber \\
 & \, + \frac{1}{2} \left( \gamma_- - a_6 \right) \left( e^{-2 \sigma} W_\beta{}^\alpha \omegared_{[\mu\alpha}{}^\beta W_{\nu]} - e^{2 \sigma} V_\beta{}^\alpha \omegared_{[\mu\alpha}{}^\beta V_{\nu]} - \frac{1}{2} W^{\lambda\rho} h_{\lambda\rho[\mu} V_{\nu]} + \right. \nonumber \\
 & \, \qquad \qquad \qquad \qquad \left. + \frac{1}{2} V^{\lambda\rho} h_{\lambda\rho[\mu} W_{\nu]} - 2 e^{-2 \sigma} W_{[\mu} W_{\nu]\rho} \dred^\rho \sigma - 2 e^{2 \sigma} V_{[\mu} V_{\nu]\rho} \dred^\rho \sigma \right) ~ .
\end{align}
\endgroup
Setting $\alpha'\to 0$ they reduce to the Buscher rules that in terms of these fields read simply as $\sigma\to -\sigma$ and $V\leftrightarrow W$. Here $\dred$ denotes the covariant derivative with respect to the reduced metric $g_{\mu\nu}$, and $\omegared_{\mu\alpha}{}^\beta$ is the reduced spin-connection. We have also defined $V_{\mu\nu}=\partial_\mu V_\nu-\partial_\nu V_\mu$, $W_{\mu\nu}=\partial_\mu W_\nu-\partial_\nu W_\mu$ and $h_{\mu\nu\rho}=3(\partial_{[\mu}b_{\nu\rho]}-\tfrac12 W_{[\mu\nu}V_{\rho]}-\tfrac12 V_{[\mu\nu}W_{\rho]})=H_{\mu\nu\rho}-3W_{[\mu\nu}V_{\rho]}$. Apart from the order-$\alpha'$ parameters $\gamma_\pm$ needed to interpolate between the bosonic and the heterotic strings (see appendix~\ref{sec:schemes}), the T-duality rules depend on coefficients $a_i,b_i,c_i$ (that are also of order $\alpha'$) so that they are valid for any scheme related to the one of BR by these field redefinitions
\begin{align}
G_{mn} & = G^{(\text{BR})}_{mn} - a_1 R_{mn} - a_2 H^2_{mn} - a_3 \nabla_m \Phi \nabla_n \Phi - a_4 \nabla_m \nabla_n \Phi - a_5 \omega_{mb}{}^a \omega_{na}{}^b - a_6 \omega_{(m}{}^{ab} H_{n)ab} ~ , \nonumber\\
B_{mn} & = B^{(\text{BR})}_{mn} - b_1 \nabla^p H_{mnp} - b_2  H_{mnp} \nabla^p \Phi - b_3 \omega_{[m}{}^{ab} H_{n]ab} ~ , \\
\Phi & = \Phi^{(\text{BR})} - c_1 R - c_2 H^2 - c_3 \nabla_p \Phi \nabla^p \Phi - c_4 \nabla^2 \Phi ~ .\nonumber
\end{align}
By turning on these coefficients we can cover all schemes typically considered in the literature, see appendix~\ref{sec:schemes} for the field redefinitions  relating them.\footnote{Writing the rules for generic $a_i,b_i,c_i$ coefficients as above, or in other words translating them into  new schemes starting from a given one, is straightforward although it requires work to compute all tensors in the dimensional reduction. After that is done we can start from scheme $A$ where $\hat \sigma^{(A)}=-\sigma^{(A)}+\alpha'\xi$, for some $\xi$. To obtain the rules in scheme $B$ related as $\sigma^{(B)}=\sigma^{(A)}+\alpha' s$ for some $s$, we just have to compute $\hat \sigma^{(B)}=\hat\sigma^{(A)}+\alpha'\hat s=-\sigma^{(A)}+\alpha'(\xi+\hat s)=\sigma^{(B)}+\alpha'(\xi+\hat s+s)$. Notice that the fields themselves may have some explicit $\alpha'$-dependence. In this example the field $\sigma$ is odd under Buscher rules, and then the shift in the corrections $\hat s+s$ is even. Fields even under Buscher receive corrections that are odd.}

As expected, it is possible to tune the coefficients in order to set to zero all corrections to the T-duality transformations. For generic $\gamma_\pm$ it is enough to set
\be
\begin{aligned}
&a_2 =-\frac{a_1}{4}+\frac{\gamma_+}{4}\,,\ 
&&a_3=0\,,\ 
&&&a_4= 2a_1\,,\ 
a_5= -\gamma_+\,,
&&&&a_6=\gamma_-\,,\ 
&&&&&b_1=-\frac{a_1}{2}\,,\\
&b_2= a_1\,,\ 
&&b_3= 0\,,\ 
&&&c_2=-\frac{a_1}{24}-\frac{c_1}{12}\,,\ 
&&&&c_3= a_1-4c_1\,,\ 
&&&&&c_4= 4 c_1-\frac{a_1}{2}
\end{aligned}
\ee
and  T-duality reduces to the Buscher rules even to 2 loops. We will denote the fields in this (gauge-fixed) scheme by $G', B',\Phi'$. When specifying to the bosonic string ($\gamma_+=\alpha'/2, \gamma_-=0$), they are related to the HT scheme by\footnote{Here  we are further setting $a_1=c_1=0$. Turning on $a_1,c_1$ would introduce terms that vanish by means of 1-loop equations.}
\be\label{eq:HT-Bu}
\begin{aligned}
 G'_{mn}&=G^{(\text{HT})}_{mn}-\tfrac{1}{2}\alpha'\omega^{(-)ab}_{(m}\omega^{(+)}_{n)ab}=G^{(\text{HT})}_{mn}+\alpha'\left(-\tfrac{1}{2}\omega_{mab}\omega^{ab}_{n}+\tfrac18 H^2_{mn}\right)\,,\\
 B'_{mn}&=B^{(\text{HT})}_{mn}+\tfrac{1}{2}\alpha'\omega^{(-)ab}_{[m}\omega^{(+)}_{n]ab}=B^{(\text{HT})}_{mn}-\tfrac{1}{2}\alpha'H_{ab[m}\omega^{ab}_{n]}\,,\\
 \Phi'&=\Phi^{(\text{HT})}+\alpha'\tfrac{1+3q}{24}H^2\,.\\
\end{aligned}
\ee
This matches with the field redefinitions that we would write for $\bar G,\bar B,\bar \Phi$ as expected. The difference is that here we are also  imposing  the specific gauge~\eqref{eq:choice-E} and for that reason we denote the fields differently.

The rules above can be compared to the ones first derived by Kaloper and Meissner in~\cite{Kaloper:1997ux} for the bosonic string ($\gamma_+ = \alpha'/2$, $\gamma_- = 0$). The scheme used is obtained setting the coefficients to
\begin{equation}
\quad a_1 = \alpha' \quad a_2 = - \frac{\alpha'}{4} ~ , \quad b_1 = - \frac{\alpha'}{2} ~ , \quad b_3 = \frac{\alpha'}{2}  ~, \quad c_1 = \frac{\alpha'}{8} \quad c_2 = - \frac{5 \alpha'}{96} \quad c_3 = - \frac{\alpha'}{2} ~ ,
\end{equation}
and the rest of them equal to zero. To match results, one has to take into account the possibility of transforming the reduced fields by doing diffeomorphisms and gauge transformations. Under such symmetries, the T-dual reduced fields transform as:
\begin{align}
\hat{V} \to & \; \hat{V} + \alpha' \left( \mathcal{L}_{\xi} W + d v \right) ~ , \\
\hat{W} \to & \; \hat{W} + \alpha' \left( \mathcal{L}_{\xi} V + d w \right) ~ , \\
\hat{b} \to & \; \hat{b} + \alpha' \left( \mathcal{L}_{\xi} b + d \beta + \frac{1}{2} V \wedge d v + \frac{1}{2} W \wedge d w \right)  ~ ,
\end{align}
and the remaining fields transform normally under diffemorphisms. We are restricting to transformations which are first order in $\alpha'$, both for diffeomorphims and gauge transformations. The $d w$ and $d \beta$ terms come from gauge transformations of the $B$ field with parameter $\beta_\mu d x^{\mu} + w \, d y$, while $v$ appears when including diffeomorphisms of the form $y \to y + \alpha' v$. Choosing the following set of parameters
\begin{equation}
\xi^{\mu} = \dred^{\mu} \sigma ~, \quad w = - V_{\nu} \dred^{\nu} \sigma ~ , \quad v = - W_{\nu} \dred^{\nu} \sigma ~, \quad \beta_{\mu} = \left( b_{\mu \nu} - \frac{1}{2} V_{\mu} W_{\nu} - \frac{1}{2} W_{\mu} V_{\nu} \right) \dred^{\nu} \sigma ~ ,
\end{equation}
we obtain the following set of rules
\begin{align}
\hat{\sigma} = & \, - \sigma + \frac{\alpha'}{2} \left[ \frac{e^{2 \sigma}}{4} V_{\lambda \rho} V^{\lambda \rho} + \frac{e^{-2 \sigma}}{4} W_{\lambda \rho} W^{\lambda \rho} + 2 \left( \dred \sigma \right)^2  \right] ~ , \\
\hat{V}_{\mu} = & \, W_{\mu} + \frac{\alpha'}{2} \left[ \frac{e^{2 \sigma}}{2} h_{\mu \lambda \rho} V^{\lambda \rho} + 2 W_{\mu \rho} \dred^\rho \sigma  \right] ~ , \\
\hat{W}_{\mu} = & \, V_{\mu} - \frac{\alpha'}{2} \left[ \frac{e^{- 2 \sigma}}{2} h_{\mu \lambda \rho} W^{\lambda \rho} - 2 V_{\mu \rho} \dred^\rho \sigma  \right] ~ , \\
\hat{b}_{\mu \nu} = & \, b_{\mu \nu} + \alpha' \left[ V_{[\mu}{}^\rho W_{\nu]\rho} - \left( V_{[\mu} W_{\nu]\rho} + W_{[\mu} V_{\nu]\rho} \right) \dred^\rho \sigma \right. \nonumber \\
 & \qquad \qquad \quad \left. - \frac{e^{2 \sigma}}{4} V_{[\mu} h_{\nu] \lambda \rho} V^{\lambda \rho} + \frac{e^{-2 \sigma}}{4} W_{[\mu} h_{\nu] \lambda \rho} W^{\lambda \rho} \right] ~ ,
\end{align}
and both $g_{\mu \nu}$ and $\phi$ remain invariant. These match with the rules given by Kaloper and Meissner in~\cite{Kaloper:1997ux} up to the sign of the $\alpha'$ correction of the $b$ field.\footnote{
The fact that this is a typo in~\cite{Kaloper:1997ux} is confirmed by the fact that there (4.9) and (4.11) are not compatible. For the  field $H$ of~\cite{Kaloper:1997ux} (here $h$) which is even under T-duality at leading order in $\alpha'$, the correction to the T-duality transformation should rather be $-2$ the expression in (4.9). For odd fields the same contribution would be instead multiplied by $+2$. This easily follows from  the first calculation they do to remove by a field redefinition the part of the action that is odd under T-duality, which is later reinterpreted as a correction to the T-duality transformation. Since the expressions in~\cite{Garousi:2019wgz} agree with those in~\cite{Kaloper:1997ux} we disagree also with that paper.}

Diffeomorphisms and gauge transformations of the reduced fields can also be used to simplify the rules and to obtain some nice expressions for the T-duality rules without the need of the dimensional reduction. We do this in a Lorentz-covariant scheme, the HT scheme for the bosonic string introduced in \eqref{eq:HT-Bu} where we fix $q = -1/3$. Using the same parameters for the transformations presented in the previous paragraph, it is possible to obtain the following rules for the T-duality transformation\footnote{Here 1-loop equations of motion were used to simplify the form of the corrections.}
\begin{align}
\hat{M}_{yy} = & \, \frac{1}{M_{yy}} ~ , \quad \hat{M}_{y\mu} = \, \frac{M_{y\mu}}{M_{yy}} ~ , \quad \hat{M}_{\mu y} = \, - \frac{M_{\mu y}}{M_{yy}} ~ , \\
\hat{M}_{\mu \nu} = & \, M_{\mu \nu} - \frac{M_{\mu y} M_{y \nu}}{M_{yy}} - \alpha' \left[ \frac{1}{M_{yy}} R^{(-)}_{\mu y \nu y} - \frac{1}{\hat{M}_{yy}} \hat{R}^{(-)}_{\mu y \nu y} \right] ~ ,\\
\hat{\Phi} = & \, \Phi - \frac{1}{2} \log M_{yy} - \frac{\alpha'}{8} \left[ R^{(-)} - \hat{R}^{(-)} \right] ~ .
\end{align}
In these expressions 
{
\be
R^{(-)}_{mna}{}^b=2\partial_{[m}\omega_{n]a}^{(-)b}+2\omega_{[ma}^{(-)c}\omega_{n]c}^{(-)b}
\ee
}
is the Riemann tensor constructed from the torsionful connection $\omega^{(-)}$, {$R^{(-)} $ the corresponding Ricci scalar} and $M_{m n} = G_{mn} - B_{m n}$. Note also that the dual appears explicitly in the $\alpha'$ corrections but, to the order needed, it can be calculated using just the standard Buscher rules.

\subsection{Corrections to TsT transformations}
The fact that there exists a (gauge-fixed) scheme --- for the sake of the discussion we will call it the ``Buscher scheme'' --- such that T-duality is given  just by the Buscher rules is useful. Here we use it to obtain an expression for $\alpha'$-corrections to TsT transformations that does not necessarily use all the knowledge of DFT. TsT transformations are a special case of YB deformations, and we will show that the result agrees with that  in section~\ref{sec:alpha-prime}.

In order to do the TsT transformation we assume that there are two $U(1)$ isometries with corresponding coordinates $y_1$ and $y_2$, and to avoid burdening the notation we will continue labelling by $x^\mu$ the rest of the coordinates.\footnote{The reader should be careful, since when doing T-duality along $y_1$ the coordinate $y_2$ should be treated on the same footing as $x^\mu$ when using the T-duality rules~\eqref{rulesGeneric}.} We will do a T-duality $y_1\to T(y_1)$ followed by a shift $y_2\to y_2-\eta T(y_1)$ and by another T-duality $T(y_1)\to y_1$. TsT transformations are special cases of YB if we take $\Theta=\eta k_1\wedge k_2$, where $k_i=\partial_{y_i}$ are Killing vectors. Each step will be performed in the scheme that is most convenient. Therefore, when doing T-duality we will prefer to move to the Buscher scheme, while when doing the shift we will prefer to go to a covariant scheme. We will show that the $\alpha'$-corrections to TsT transformations can be understood as arising from these shifts coming from the scheme changes. Because these scheme-changing shifts arise at intermediate steps, we will have to look at how they are further modified by the remaining steps in the TsT transformation.

Suppose we start from the HT scheme. In order to do the first T-duality on $y_1$ we find convenient to first go to the Buscher scheme. This is achieved by implementing the redefinitions~\eqref{eq:HT-Bu} after taking care of choosing the vielbein as in~\eqref{eq:choice-E}. This effectively shifts the fields at order $\alpha'$ as $\delta_1(G_{mn}-B_{mn})=-\tfrac12 \omega^{(-)}_{mab}\omega^{(+)ab}_{n}$. We can immediately account for this contribution in the final result: because we will have to do a TsT transformation including this contribution $\delta_1$ (and we only care about the order $\alpha'$) we are essentially shifting the original metric and $B$-field as $G-B\to G-B + \delta_1(G-B)$ appearing in the map~\eqref{eq:SW}. After expanding to first order in $\alpha'$ we obtain the first contribution to the $\alpha'$ correction of the final result
\be
-\tfrac12[(1+ (G-B)\Theta)^{-1}]_m{}^p \omega^{(-)}_{pab}\omega^{(+)ab}_{q}[(1+\Theta (G-B))^{-1}]^q{}_n\,.
\ee
While in the Buscher scheme we can easily do the first T-duality on $y_1$ because we just need to use the Buscher rules. Notice that under Buscher the gauge choice~\eqref{eq:choice-E} is preserved. 

To perform the shift it is more convenient to go back to the HT scheme, which is covariant. That means that we will have to use~\eqref{eq:HT-Bu} again, although now it will be done using the data of the T-dual background $\delta_2(\hat G_{mn}-\hat B_{mn})=+\tfrac12 \hat\omega^{(-)}_{mab}\hat\omega^{(+)ab}_{n}$. A hat is used to denote that the first T-dualization has already been done. Notice that under the first T-duality and shift, the vielbein $e_m{}^a$  (in matrix form) changes as
\be
\left(
\begin{array}{ccc}
 e^{\sigma }  &  0 &  0   \\
e^{\sigma } V_{y_2}  & \ered_{y_2}{}^2 & \ered_{y_2}{}^\alpha \\
e^{\sigma } V_\mu & \ered_\mu{}^2 & \ered_\mu{}^\alpha \\
\end{array}
\right)
\xrightarrow{T}
\left(
\begin{array}{ccc}
 e^{-\sigma }  &  0 &  0   \\
e^{-\sigma } W_{y_2}  & \ered_{y_2}{}^2 & \ered_{y_2}{}^\alpha \\
e^{-\sigma } W_\mu & \ered_\mu{}^2 & \ered_\mu{}^\alpha \\
\end{array}
\right)
\xrightarrow{s}
\left(
\begin{array}{ccc}
 e^{-\sigma } (1- \eta W_{y_2}) &  -\eta \ered_{y_2}{}^2 &  -\eta  \ered_{y_2}{}^\alpha   \\
e^{-\sigma } W_{y_2}  & \ered_{y_2}{}^2 & \ered_{y_2}{}^\alpha \\
e^{-\sigma } W_\mu & \ered_\mu{}^2 & \ered_\mu{}^\alpha \\
\end{array}
\right).
\ee
The shift is spoiling the choice~\eqref{eq:choice-E} for the vielbein, and that is an important point because we will want to restore this gauge before going back to the Buscher scheme and implementing the last T-duality. To achieve it we implement the Lorentz transformation $e_m{}^a\to e_m{}^bL_b{}^a$ where
\be
L_b{}^a=\left(
\begin{array}{ccc}
 \frac{1-\eta W_{y_2}}{D^{1/2}} & \frac{ \eta e^{\sigma }
   \sqrt{g_{y_2y_2}}
   }{D^{1/2}} & 0 \\
- \frac{ \eta e^{\sigma }
   \sqrt{g_{y_2y_2}}
   }{D^{1/2}} & \frac{1- \eta W_{y_2}}{D^{1/2}} & 0 \\
 0 & 0 &
   \delta_b{}^a \\
\end{array}
\right),\qquad
\text{where }D=1-\eta W_{y_2}(2-\eta W_{y_2})+\eta^2 e^{2\sigma}g_{y_2y_2}.
\ee 
At this point one wants to go to the Buscher scheme, in order to perform the last T-duality, which will produce a new correction $\delta_3(\hat{\hat G}_{mn}-\hat{ \hat B}_{mn})=-\tfrac12 \hat {\hat\omega}^{(-)}_{mab}\hat {\hat\omega}^{(+)ab}_{n}$. Now a double hat is used to denote that a T-duality and a shift (followed by the compensating Lorentz transformation) have been implemented. The contribution $\delta_2$ (on which we implement the effect of the shift) and $\delta_3$ can be considered together. In fact all expressions from covariant terms cancel out and we are left with 
\be\label{eq:expr}
\tfrac12 \left(-\hat{\hat \omega}^{(-)}_{mab}(L^{-1}\partial_{n}L)^{ab}-\hat{\hat \omega}^{(+)}_{nab}(L^{-1}\partial_{m}L)^{ab}
+ (L^{-1}\partial_mL)_{ab}(L^{-1}\partial_nL)^{ab}\right).
\ee
In order to account for the effect of the last T-duality on the above expression one uses: the fact that in the first two terms only $(mn)\neq (y_iy_j)$ contribute, that in the summation of $a,b$ only $1,2$ contribute, the fact that under T-duality
\be
\hat \omega^{(\pm)}_{\iota \iota \beta}=-\omega^{(\pm)}_{\iota \iota \beta},\qquad
\hat \omega^{(\pm)}_{\alpha\iota \beta}=\pm \omega^{(\pm)}_{\alpha\iota \beta},\qquad
\hat \omega^{(\pm)}_{\iota \alpha\beta}=\mp \omega^{(\pm)}_{\iota \alpha\beta},\qquad
\hat \omega^{(\pm)}_{\alpha\beta\gamma}=\omega^{(\pm)}_{\alpha\beta\gamma},
\ee
and finally that the last term in~\eqref{eq:expr} vanishes if $m$ or $n$ are $y_i$, so that it actually remains the same after T-duality. After taking everything into account the result after the T-duality is simply
\be
\tfrac12 \left(\tilde{ \omega}^{(-)}_{mab}(L^{-1}\partial_{n}L)^{ab}-\tilde{ \omega}^{(+)}_{nab}(L^{-1}\partial_{m}L)^{ab}
+ (L^{-1}\partial_mL)_{ab}(L^{-1}\partial_nL)^{ab}\right).
\ee
A tilde denotes the quantities of the TsT-transformed background.

After the last T-duality has been performed, we go back from Buscher to the HT scheme using~\eqref{eq:HT-Bu} obtaining the final contribution to the $\alpha'$ corrections which is $\delta_4(\tilde{ G}_{mn}-\tilde{  B}_{mn})=+\tfrac12 \tilde {\omega}^{(-)}_{mab}\tilde {\omega}^{(+)ab}_{n}$.

Collecting together all contributions we obtain the $\alpha'$ correction to the TsT deformed background in the HT scheme
\be
\begin{aligned}
\delta(\tilde G-\tilde B)_{mn}=&\tfrac12 \left(\tilde{ \omega}^{(-)}_{mab}-(L^{-1}\partial_{m}L)_{ab}\right)\left(\tilde{ \omega}^{(+)ab}_{n}+(L^{-1}\partial_{n}L)^{ab}\right)
+ (L^{-1}\partial_mL)_{ab}(L^{-1}\partial_nL)^{ab}\\
&-\tfrac12[(1+ (G-B)\Theta)^{-1}]_m{}^p \omega^{(-)}_{pab}\omega^{(+)ab}_{q}[(1+\Theta (G-B))^{-1}]^q{}_n.
\end{aligned}
\ee
Because of the steps of TsT, the vielbein used to construct the above spin-connection of the deformed model is defined as $\tilde e_a{}^m = L_a{}^be_b{}^n(1- (G+B)\Theta)_n{}^m$, where the undeformed vielbein must respect~\eqref{eq:choice-E}, and one can check that the Lorentz transformation used here is related to the one in~\eqref{eq:Lambda-} simply as $L^{2}=\tilde \Lambda$. To compare to the result~\eqref{eq:non-cov} we need to use the same deformed vielbein used there, meaning that we should rather take
$\tilde e_a{}^m = (L^{2})_a{}^be_b{}^n(1- (G+B)\Theta)_n{}^m$. After taking into account this extra Lorentz transformation we match with~\eqref{eq:non-cov} in the case of TsT if we remember that $B^{\text{WZW}}$ can be taken to be zero, and if we use that we for TsT we can write $L^{-1}dL=dLL^{-1}$ because here $L$ is essentially a $2\times 2$ anti-symmetric matrix and it commutes with itself.

With a similar reasoning we can obtain the $\alpha'$-corrections to the dilaton of the TsT-transformed background. The simplification in this case is that the dilaton is insensitive to the shift, because by assumption it is isometric and the field redefinitions for the dilaton between the schemes of Buscher and HT are covariant. In HT scheme at generic $q$ we get
\be
\begin{aligned}
\widetilde \Phi&=\Phi+\frac{1}{2}\log\frac{\widetilde G_{y_1y_1}}{G_{y_1y_1}}+\alpha'\left[\frac{1+3q}{24}(H^2-\widetilde H^2)-\frac12 \left(\frac{\delta_1 G_{y_1y_1}}{G_{y_1y_1}}+\frac{\delta_4 G_{y_1y_1}}{\widetilde G_{y_1y_1}}\right)\right].
\end{aligned}
\ee
At $q=1/6$ on finds
\be
e^{-2\widetilde\Phi}\sqrt{-\det \widetilde G}=
e^{-2\Phi}\sqrt{-\det  G}\,,
\ee
which is in agreement with~\eqref{eq:dil}, {since there the result was written when setting $q=0$, and one therefore has the extra $H^2$-terms.}

\section{Concluding comments}
{In this paper we have demonstrated that it is possible to extend the YB deformation as a solution-generating technique in string theory at least to first order in the $\alpha'$-expansion. The explicit expression that we found for the corrections allowed us to test successfully our results on explicit examples. We expect our formula to be useful when addressing specific questions on the $\alpha'$-corrected YB-deformed backgrounds. For example, it would be interesting to see whether the singularities that are sometimes introduced by the deformation procedure are in fact cured by $\alpha'$-corrections. Another point is the computation of physical observables on the deformed backgrounds --- such as entropy calculations in black hole solutions,\footnote{{While in this paper we have considered only the case of the bosonic string, it is easy to generalize our results to generic values of the parameters $a,b$ interpolating between the bosonic and the heterotic string.}} see e.g.~\cite{Arvanitakis:2016zes,Cano:2018qev,Edelstein:2019wzg,Elgood:2020xwu} --- for which the explicit corrections are needed. 
It would be also interesting to investigate the relation to (quantum) integrability when considering YB-deformations of integrable 2-dimensional $\sigma$-models. }

We have seen that the $\alpha'$-correction to YB deformations comes from a compensating Lorentz transformation under which the $O(d,d)$ covariant metric and $B$-field transform anomalously. It is natural to expect that the same should be true also for T-duality. In fact abelian and non-abelian T-dualities are used to construct the YB deformation and they can also be obtained as a limit (sending the deformation parameter to infinity) of YB deformations. In fact we have already argued that for abelian T-duality the correction is given by precisely the same mechanism. It is therefore very natural to expect the first $\alpha'$-correction to non-abelian T-duality\footnote{Using very different arguments NATD has been argued to preserve Weyl invariance at least to 2 loops, and probably to all orders in \cite{Hoare:2019ark,Hoare:2019mcc}.} (on a unimodular algebra) to be given by the same expression, with the Lorentz transformation required for NATD substituted for $\tilde\Lambda$ in (\ref{eq:non-cov}) and (\ref{eq:dil}).

As in previous works on YB deformations and NATD (see e.g.~\cite{Hoare:2016wsk,Borsato:2016pas,Borsato:2018idb}) here it was assumed that the undeformed $B$-field and vielbein are isometric, i.e. that they have vanishing Lie derivative with respect to the Killing vectors entering $\Theta$. {The covariant form of the corrections we have found seems to be valid more generally but it would be interesting to analyze more systematically how to relax these assumptions.}

In~\cite{Borsato:2018spz} YB deformations of strings on $AdS_3\times S^3$ were studied, and their relation to marginal deformations of WZW models was analyzed. The results of the current paper show that marginal deformations of current algebras include (at least to 2 loops and probably to all loops) also cases which do not solve the ``strong version'' of the marginality condition of Chaudhuri and Schwartz~\cite{Chaudhuri:1988qb}, see~\cite{Borsato:2018spz} for more details. These additional possibilities arise when considering algebras that are not compact. 
Let us also comment that the deformation generated by the unimodular non-abelian $R_9$ of~\cite{Borsato:2018spz} must be marginal to all loops, since it can be simply understood as a non-commuting sequence of TsT transformations.
 
{We expect that generalizations of our discussion to a construction in the spirit of the $\mathcal E$-model of Klim\v{c}ik~\cite{Klimcik:1995dy,Stern:1998my,Klimcik:2015gba} will lead to an understanding of the form of $\alpha'$-corrections for the $\eta$-deformation~\cite{Klimcik:2008eq,Delduc:2013qra}, the $\lambda$-deformation~\cite{Sfetsos:2013wia,Hollowood:2014qma}, and to Poisson-Lie T-duality~\cite{Klimcik:1995ux}.}

{Another important question we hope to return to is if the structure of the correction found here persists beyond first order in $\alpha'$ or whether novel corrections are required at order $\alpha'^2$.}

\section*{Acknowledgements}
We thank D. Marqu\'es for discussions and A. Tseytlin for comments on the manuscript. The work of RB is supported by the fellowship of ``la Caixa Foundation'' (ID 100010434) with code LCF/BQ/PI19/11690019, by FPA2017-84436-P, and by Xunta de Galicia (ED431C 2017/07). AVL is supported by the Spanish MECD fellowship FPU16/06675. RB and AVL are also supported by the European Regional Development Fund (ERDF/FEDER program), by the ``Mar\'ia de Maeztu'' Units of Excellence program MDM-2016-0692 and by the Spanish Research State Agency. The work of LW is supported by the grant ``Integrable Deformations'' (GA20-04800S) from the Czech Science Foundation (GACR).

\appendix

\section{Field redefinitions between different schemes}\label{sec:schemes}
In this appendix  we collect the field redefinitions needed to relate --- to first order in the $\alpha'$ expansion --- the schemes we use in this paper and others relevant in the literature. These are the schemes of Hull and Townsend (HT)~\cite{Hull:1987yi}, Metsaev and Tseytlin (MT)~\cite{Metsaev:1987zx}, Kaloper and Meissner (KP)~\cite{Meissner:1996sa,Kaloper:1997ux}, Bergshoeff and de Roo (BR)~\cite{Bergshoeff:1988nn,Bergshoeff:1989de}.
From~\cite{Marques:2015vua} we read
\be
\begin{aligned}
G^{(\text{BR})}_{mn}&=G^{(\text{MT})}_{mn}-\tfrac12 \gamma_+ H^2_{mn},\\
B^{(\text{BR})}_{mn}&=B^{(\text{MT})}_{mn}-\gamma_+ (\nabla^pH_{mnp}-2H_{mnp}\nabla^p\Phi+H_{[m}{}^{ab}\omega_{n]ab})\\
&\simeq B^{(\text{MT})}_{mn}-\gamma_+ H_{[m}{}^{ab}\omega_{n]ab},\\
\Phi^{(\text{BR})}&=\Phi^{(\text{MT})}-\tfrac{1}{8}\gamma_+ H^2.
\end{aligned}
\ee
The symbol $\simeq$ is used when the expressions are simplified by means of the 1-loop equations of motion.
We relate the parameters $\gamma_\pm=\mp(a\pm b)/4$ to $a,b$ used in~\cite{Marques:2015vua}. The bosonic string is obtained at $\gamma_+=\alpha'/2, \gamma_-=0$ and the heterotic string at $\gamma_\pm=\pm \alpha'/4$. In the following we will specify to the case of the bosonic string.
To relate HT and MT schemes we use
\be
\begin{aligned}
&G^{(\text{HT})}_{mn}=G^{(\text{MT})}_{mn}-\tfrac{1}{2}\alpha'H^2_{mn},\\
&B^{(\text{HT})}_{mn}=B^{(\text{MT})}_{mn},\\
&\Phi^{(\text{HT})}=\Phi^{(\text{MT})}+\tfrac{1}{8}\alpha'(-1+\tfrac16(1-6q))H^2.\\
\end{aligned}
\ee
The parameter $q$ appears in~\cite{Hull:1987yi}, and we normally set $q=0$ in the rest of the paper as in~\cite{Borsato:2019oip}. Notice that the sign of the correction to the metric differs from what one would read in~\cite{Hull:1987yi}. We have checked that this is the correct sign in order to have the correct $\alpha'$-corrections for T-duality and YB deformations.
From~\cite{Meissner:1996sa} we read that 
\be
\begin{aligned}
&G^{(\text{MT})}_{mn}=G^{(\text{KM})}_{mn}+\alpha'R_{mn},\\
&B^{(\text{MT})}_{mn}=B^{(\text{KM})}_{mn}-\alpha'H_{mnp}\nabla^P\Phi,\\
&\Phi^{(\text{MT})}=\Phi^{(\text{KM})}+\alpha'(\tfrac18 R-\tfrac12 (\partial\Phi)^2+\tfrac{1}{96}H^2).\\
\end{aligned}
\ee
The fields of the non-covariant scheme that follows from the DFT formulation are denoted simply with a bar $\bar G,\bar B,\bar \Phi$. They are related to the fields in the HT scheme as
\be
\begin{aligned}
\bar G_{mn}&=G^{(\text{HT})}_{mn}-\tfrac{1}{2}\alpha'\omega^{(-)ab}_{(m}\omega^{(+)}_{n)ab}=G^{(\text{HT})}_{mn}+\alpha'\left(-\tfrac{1}{2}\omega_{mab}\omega^{ab}_{n}+\tfrac18 H^2_{mn}\right),\\
\bar B_{mn}&=B^{(\text{HT})}_{mn}+\tfrac{1}{2}\alpha'\omega^{(-)ab}_{[m}\omega^{(+)}_{n]ab}=B^{(\text{HT})}_{mn}-\tfrac{1}{2}\alpha'H_{ab[m}\omega^{ab}_{n]}.\\
\end{aligned}
\ee

\bibliographystyle{nb}
\bibliography{biblio}{}

\begin{thebibliography}{10}
\ifx\href\asklfhas\newcommand{\href}[2]{#2}\fi
\ifx\arxivref\asklfhas\newcommand{\arxivref}[2]{\href{http://arxiv.org/abs/#1}{#2}}\fi
\ifx\doiref\asklfhas\newcommand{\doiref}[2]{\href{http://dx.doi.org/#1}{#2}}\fi
\raggedright
\small
\parskip 0pt

\bibitem{Klimcik:2002zj}
C.~Klimcik,
\textit{``{Yang-Baxter sigma models and dS/AdS T duality}''},
\textsf{\doiref{10.1088/1126-6708/2002/12/051}{JHEP~0212,~051~(2002)}},
\texttt{\arxivref{hep-th/0210095}{hep-th/0210095}}.

\bibitem{Klimcik:2008eq}
C.~Klimcik,
\textit{``{On integrability of the Yang-Baxter sigma-model}''},
\textsf{\doiref{10.1063/1.3116242}{J.Math.Phys.~50,~043508~(2009)}},
\texttt{\arxivref{0802.3518}{arxiv:0802.3518}}.

\bibitem{Delduc:2013qra}
F.~Delduc, M.~Magro and B.~Vicedo,
\textit{``{An integrable deformation of the AdS$_5 \times$S$^5$ superstring
  action}''},
\textsf{\doiref{10.1103/PhysRevLett.112.051601}{Phys.Rev.Lett.~112,~051601~(2014)}},
\texttt{\arxivref{1309.5850}{arxiv:1309.5850}}.

\bibitem{Kawaguchi:2014qwa}
I.~Kawaguchi, T.~Matsumoto and K.~Yoshida,
\textit{``{Jordanian deformations of the $AdS_5 x S^5$ superstring}''},
\textsf{\doiref{10.1007/JHEP04(2014)153}{JHEP~1404,~153~(2014)}},
\texttt{\arxivref{1401.4855}{arxiv:1401.4855}}.

\bibitem{Hoare:2016wsk}
B.~Hoare and A.~A.~Tseytlin,
\textit{``{Homogeneous Yang-Baxter deformations as non-abelian duals of the
  AdS$_5$ sigma-model}''},
\textsf{\doiref{10.1088/1751-8113/49/49/494001}{J.~Phys.~A49,~494001~(2016)}},
\texttt{\arxivref{1609.02550}{arxiv:1609.02550}}.

\bibitem{Borsato:2016pas}
R.~Borsato and L.~Wulff,
\textit{``{Integrable Deformations of $T$-Dual $\sigma$ Models}''},
\textsf{\doiref{10.1103/PhysRevLett.117.251602}{Phys.~Rev.~Lett.~117,~251602~(2016)}},
\texttt{\arxivref{1609.09834}{arxiv:1609.09834}}.

\bibitem{Borsato:2018idb}
R.~Borsato and L.~Wulff,
\textit{``{Non-abelian T-duality and Yang-Baxter deformations of Green-Schwarz
  strings}''},
\textsf{\doiref{10.1007/JHEP08(2018)027}{JHEP~1808,~027~(2018)}},
\texttt{\arxivref{1806.04083}{arxiv:1806.04083}}.

\bibitem{Osten:2016dvf}
D.~Osten and S.~J.~van~Tongeren,
\textit{``{Abelian Yang–Baxter deformations and TsT transformations}''},
\textsf{\doiref{10.1016/j.nuclphysb.2016.12.007}{Nucl.~Phys.~B915,~184~(2017)}},
\texttt{\arxivref{1608.08504}{arxiv:1608.08504}}.

\bibitem{Alvarez:1994np}
E.~Alvarez, L.~Alvarez-Gaume and Y.~Lozano,
\textit{``{On nonAbelian duality}''},
\textsf{\doiref{10.1016/0550-3213(94)90093-0}{Nucl.~Phys.~B424,~155~(1994)}},
\texttt{\arxivref{hep-th/9403155}{hep-th/9403155}}.

\bibitem{Elitzur:1994ri}
S.~Elitzur, A.~Giveon, E.~Rabinovici, A.~Schwimmer and G.~Veneziano,
\textit{``{Remarks on nonAbelian duality}''},
\textsf{\doiref{10.1016/0550-3213(94)00426-F}{Nucl.~Phys.~B435,~147~(1995)}},
\texttt{\arxivref{hep-th/9409011}{hep-th/9409011}}.

\bibitem{Arutyunov:2015mqj}
G.~Arutyunov, S.~Frolov, B.~Hoare, R.~Roiban and A.~A.~Tseytlin,
\textit{``{Scale invariance of the $\eta$-deformed $AdS_5\times S^5$
  superstring, T-duality and modified type II equations}''},
\textsf{\doiref{10.1016/j.nuclphysb.2015.12.012}{Nucl.~Phys.~B903,~262~(2016)}},
\texttt{\arxivref{1511.05795}{arxiv:1511.05795}}.

\bibitem{Wulff:2016tju}
L.~Wulff and A.~A.~Tseytlin,
\textit{``{Kappa-symmetry of superstring sigma model and generalized 10d
  supergravity equations}''},
\textsf{\doiref{10.1007/JHEP06(2016)174}{JHEP~1606,~174~(2016)}},
\texttt{\arxivref{1605.04884}{arxiv:1605.04884}}.

\bibitem{Borsato:2016ose}
R.~Borsato and L.~Wulff,
\textit{``{Target space supergeometry of $\eta$ and $\lambda$-deformed
  strings}''},
\textsf{\doiref{10.1007/JHEP10(2016)045}{JHEP~1610,~045~(2016)}},
\texttt{\arxivref{1608.03570}{arxiv:1608.03570}}.

\bibitem{Sakamoto:2018krs}
J.-I.~Sakamoto and Y.~Sakatani,
\textit{``{Local $\beta$-deformations and Yang-Baxter sigma model}''},
\textsf{\doiref{10.1007/JHEP06(2018)147}{JHEP~1806,~147~(2018)}},
\texttt{\arxivref{1803.05903}{arxiv:1803.05903}}.

\bibitem{Wulff:2018aku}
L.~Wulff,
\textit{``{Trivial solutions of generalized supergravity vs non-abelian
  T-duality anomaly}''},
\textsf{\doiref{10.1016/j.physletb.2018.04.025}{Phys.~Lett.~B781,~417~(2018)}},
\texttt{\arxivref{1803.07391}{arxiv:1803.07391}}.

\bibitem{Borsato:2018spz}
R.~Borsato and L.~Wulff,
\textit{``{Marginal deformations of WZW models and the classical Yang–Baxter
  equation}''},
\textsf{\doiref{10.1088/1751-8121/ab1b9c}{J.~Phys.~A52,~225401~(2019)}},
\texttt{\arxivref{1812.07287}{arxiv:1812.07287}}.

\bibitem{Sakamoto:2017cpu}
J.-i.~Sakamoto, Y.~Sakatani and K.~Yoshida,
\textit{``{Homogeneous Yang-Baxter deformations as generalized
  diffeomorphisms}''},
\textsf{\doiref{10.1088/1751-8121/aa8896}{J.~Phys.~A50,~415401~(2017)}},
\texttt{\arxivref{1705.07116}{arxiv:1705.07116}}.

\bibitem{Lust:2018jsx}
D.~L{\"u}st and D.~Osten,
\textit{``{Generalised fluxes, Yang-Baxter deformations and the O(d,d)
  structure of non-abelian T-duality}''},
\textsf{\doiref{10.1007/JHEP05(2018)165}{JHEP~1805,~165~(2018)}},
\texttt{\arxivref{1803.03971}{arxiv:1803.03971}}.

\bibitem{Bakhmatov:2018bvp}
I.~Bakhmatov and E.~T.~Musaev,
\textit{``{Classical Yang-Baxter equation from $\beta$-supergravity}''},
\textsf{\doiref{10.1007/JHEP01(2019)140}{JHEP~1901,~140~(2019)}},
\texttt{\arxivref{1811.09056}{arxiv:1811.09056}}.

\bibitem{Baron:2017dvb}
W.~H.~Baron, J.~J.~Fernandez-Melgarejo, D.~Marques and C.~Nunez,
\textit{``{The Odd story of $\alpha'$-corrections}''},
\textsf{\doiref{10.1007/JHEP04(2017)078}{JHEP~1704,~078~(2017)}},
\texttt{\arxivref{1702.05489}{arxiv:1702.05489}}.

\bibitem{Borsato:2019oip}
R.~Borsato and L.~Wulff,
\textit{``{Two-loop conformal invariance for Yang-Baxter deformed strings}''},
\texttt{\arxivref{1910.02011}{arxiv:1910.02011}}.

\bibitem{Tseytlin:1991wr}
A.~A.~Tseytlin,
\textit{``{Duality and dilaton}''},
\textsf{\doiref{10.1142/S021773239100186X}{Mod.~Phys.~Lett.~A6,~1721~(1991)}}.

\bibitem{Panvel:1992he}
J.~Panvel,
\textit{``{Higher order conformal invariance of string backgrounds obtained by
  O(d,d) transformations}''},
\textsf{\doiref{10.1016/0370-2693(92)91923-W}{Phys.~Lett.~B284,~50~(1992)}},
\texttt{\arxivref{hep-th/9204024}{hep-th/9204024}}.

\bibitem{Bergshoeff:1995cg}
E.~Bergshoeff, B.~Janssen and T.~Ortin,
\textit{``{Solution generating transformations and the string effective
  action}''},
\textsf{\doiref{10.1088/0264-9381/13/3/002}{Class.~Quant.~Grav.~13,~321~(1996)}},
\texttt{\arxivref{hep-th/9506156}{hep-th/9506156}}.

\bibitem{Haagensen:1997er}
P.~E.~Haagensen and K.~Olsen,
\textit{``{T duality and two loop renormalization flows}''},
\textsf{\doiref{10.1016/S0550-3213(97)00496-3}{Nucl.~Phys.~B504,~326~(1997)}},
\texttt{\arxivref{hep-th/9704157}{hep-th/9704157}}.

\bibitem{Kaloper:1997ux}
N.~Kaloper and K.~A.~Meissner,
\textit{``{Duality beyond the first loop}''},
\textsf{\doiref{10.1103/PhysRevD.56.7940}{Phys.~Rev.~D56,~7940~(1997)}},
\texttt{\arxivref{hep-th/9705193}{hep-th/9705193}}.

\bibitem{Jack:1999av}
I.~Jack and S.~Parsons,
\textit{``{O(d,d) invariance at two loops and three loops}''},
\textsf{\doiref{10.1103/PhysRevD.62.026003}{Phys.~Rev.~D62,~026003~(2000)}},
\texttt{\arxivref{hep-th/9911064}{hep-th/9911064}}.

\bibitem{Parsons:1999ze}
S.~Parsons,
\textit{``{T duality and conformal invariance at two loops}''},
\textsf{\doiref{10.1103/PhysRevD.61.086002}{Phys.~Rev.~D61,~086002~(2000)}},
\texttt{\arxivref{hep-th/9912105}{hep-th/9912105}}.

\bibitem{Hohm:2011si}
O.~Hohm and B.~Zwiebach,
\textit{``{On the Riemann Tensor in Double Field Theory}''},
\textsf{\doiref{10.1007/JHEP05(2012)126}{JHEP~1205,~126~(2012)}},
\texttt{\arxivref{1112.5296}{arxiv:1112.5296}}.

\bibitem{Hohm:2013jaa}
O.~Hohm, W.~Siegel and B.~Zwiebach,
\textit{``{Doubled $\alpha'$-geometry}''},
\textsf{\doiref{10.1007/JHEP02(2014)065}{JHEP~1402,~065~(2014)}},
\texttt{\arxivref{1306.2970}{arxiv:1306.2970}}.

\bibitem{Hohm:2014xsa}
O.~Hohm and B.~Zwiebach,
\textit{``{Double field theory at order $\alpha'$}''},
\textsf{\doiref{10.1007/JHEP11(2014)075}{JHEP~1411,~075~(2014)}},
\texttt{\arxivref{1407.3803}{arxiv:1407.3803}}.

\bibitem{Hohm:2014eba}
O.~Hohm and B.~Zwiebach,
\textit{``{Green-Schwarz mechanism and $\alpha'$-deformed Courant brackets}''},
\textsf{\doiref{10.1007/JHEP01(2015)012}{JHEP~1501,~012~(2015)}},
\texttt{\arxivref{1407.0708}{arxiv:1407.0708}}.

\bibitem{Marques:2015vua}
D.~Marques and C.~A.~Nunez,
\textit{``{T-duality and $\alpha'$-corrections}''},
\textsf{\doiref{10.1007/JHEP10(2015)084}{JHEP~1510,~084~(2015)}},
\texttt{\arxivref{1507.00652}{arxiv:1507.00652}}.

\bibitem{Siegel:1993th}
W.~Siegel,
\textit{``{Superspace duality in low-energy superstrings}''},
\textsf{\doiref{10.1103/PhysRevD.48.2826}{Phys.~Rev.~D48,~2826~(1993)}},
\texttt{\arxivref{hep-th/9305073}{hep-th/9305073}}.

\bibitem{Siegel:1993xq}
W.~Siegel,
\textit{``{Two vierbein formalism for string inspired axionic gravity}''},
\textsf{\doiref{10.1103/PhysRevD.47.5453}{Phys.~Rev.~D47,~5453~(1993)}},
\texttt{\arxivref{hep-th/9302036}{hep-th/9302036}}.

\bibitem{Hohm:2010xe}
O.~Hohm and S.~K.~Kwak,
\textit{``{Frame-like Geometry of Double Field Theory}''},
\textsf{\doiref{10.1088/1751-8113/44/8/085404}{J.~Phys.~A44,~085404~(2011)}},
\texttt{\arxivref{1011.4101}{arxiv:1011.4101}}.

\bibitem{Aldazabal:2013sca}
G.~Aldazabal, D.~Marques and C.~Nunez,
\textit{``{Double Field Theory: A Pedagogical Review}''},
\textsf{\doiref{10.1088/0264-9381/30/16/163001}{Class.~Quant.~Grav.~30,~163001~(2013)}},
\texttt{\arxivref{1305.1907}{arxiv:1305.1907}}.

\bibitem{Hohm:2013bwa}
O.~Hohm, D.~L{\"u}st and B.~Zwiebach,
\textit{``{The Spacetime of Double Field Theory: Review, Remarks, and
  Outlook}''},
\textsf{\doiref{10.1002/prop.201300024}{Fortsch.~Phys.~61,~926~(2013)}},
\texttt{\arxivref{1309.2977}{arxiv:1309.2977}}.

\bibitem{Berman:2013eva}
D.~S.~Berman and D.~C.~Thompson,
\textit{``{Duality Symmetric String and M-Theory}''},
\textsf{\doiref{10.1016/j.physrep.2014.11.007}{Phys.~Rept.~566,~1~(2014)}},
\texttt{\arxivref{1306.2643}{arxiv:1306.2643}}.

\bibitem{Araujo:2017jkb}
T.~Araujo, I.~Bakhmatov, E.~O.~Colg\'ain, J.~Sakamoto, M.~M.~Sheikh-Jabbari and
  K.~Yoshida,
\textit{``{Yang-Baxter $\sigma$-models, conformal twists, and noncommutative
  Yang-Mills theory}''},
\textsf{\doiref{10.1103/PhysRevD.95.105006}{Phys.~Rev.~D95,~105006~(2017)}},
\texttt{\arxivref{1702.02861}{arxiv:1702.02861}}.

\bibitem{Catal-Ozer:2019tmm}
A.~Çatal~Özer and S.~Tunal\i,
\textit{``{Yang-Baxter Deformation as an O(d,d) Transformation}''},
\textsf{\doiref{10.1088/1361-6382/ab6f7e}{Class.~Quant.~Grav.~37,~075003~(2020)}},
\texttt{\arxivref{1906.09053}{arxiv:1906.09053}}.

\bibitem{Edelstein:2019wzg}
J.~D.~Edelstein, K.~Sfetsos, J.~A.~Sierra-Garcia and A.~Vilar~L\'opez,
\textit{``{T-duality equivalences beyond string theory}''},
\textsf{\doiref{10.1007/JHEP05(2019)082}{JHEP~1905,~082~(2019)}},
\texttt{\arxivref{1903.05554}{arxiv:1903.05554}}.

\bibitem{Eloy:2019hnl}
C.~Eloy, O.~Hohm and H.~Samtleben,
\textit{``{Green-Schwarz Mechanism for String Dualities}''},
\textsf{\doiref{10.1103/PhysRevLett.124.091601}{Phys.\~Rev.\~Lett.~124,~091601~(2020)}},
\texttt{\arxivref{1912.01700}{arxiv:1912.01700}}.

\bibitem{Bergshoeff:1988nn}
E.~Bergshoeff and M.~de~Roo,
\textit{``{Supersymmetric Chern-simons Terms in Ten-dimensions}''},
\textsf{\doiref{10.1016/0370-2693(89)91420-2}{Phys.~Lett.~B218,~210~(1989)}}.

\bibitem{Bergshoeff:1989de}
E.~A.~Bergshoeff and M.~de~Roo,
\textit{``{The Quartic Effective Action of the Heterotic String and
  Supersymmetry}''},
\textsf{\doiref{10.1016/0550-3213(89)90336-2}{Nucl.~Phys.~B328,~439~(1989)}}.

\bibitem{Garousi:2019wgz}
M.~R.~Garousi,
\textit{``{Four-derivative couplings via the $T$-duality invariance
  constraint}''},
\textsf{\doiref{10.1103/PhysRevD.99.126005}{Phys.~Rev.~D99,~126005~(2019)}},
\texttt{\arxivref{1904.11282}{arxiv:1904.11282}}.

\bibitem{Arvanitakis:2016zes}
A.~S.~Arvanitakis and C.~D.~A.~Blair,
\textit{``{Black hole thermodynamics, stringy dualities and double field
  theory}''},
\textsf{\doiref{10.1088/1361-6382/aa5a59}{Class.~Quant.~Grav.~34,~055001~(2017)}},
\texttt{\arxivref{1608.04734}{arxiv:1608.04734}}.

\bibitem{Cano:2018qev}
P.~A.~Cano, P.~Meessen, T.~Ort\'in and P.~F.~Ram\'irez,
\textit{``{$\alpha'$-corrected black holes in String Theory}''},
\textsf{\doiref{10.1007/JHEP05(2018)110}{JHEP~1805,~110~(2018)}},
\texttt{\arxivref{1803.01919}{arxiv:1803.01919}}.

\bibitem{Elgood:2020xwu}
Z.~Elgood and T.~Ortin,
\textit{``{T duality and Wald entropy formula in the Heterotic Superstring
  effective action at first order in $\alpha'$}''},
\texttt{\arxivref{2005.11272}{arxiv:2005.11272}}.

\bibitem{Hoare:2019ark}
B.~Hoare, N.~Levine and A.~A.~Tseytlin,
\textit{``{Integrable 2d sigma models: quantum corrections to geometry from RG
  flow}''},
\textsf{\doiref{10.1016/j.nuclphysb.2019.114798}{Nucl.~Phys.~B949,~114798~(2019)}},
\texttt{\arxivref{1907.04737}{arxiv:1907.04737}}.

\bibitem{Hoare:2019mcc}
B.~Hoare, N.~Levine and A.~A.~Tseytlin,
\textit{``{Integrable sigma models and 2-loop RG flow}''},
\textsf{\doiref{10.1007/JHEP12(2019)146}{JHEP~1912,~146~(2019)}},
\texttt{\arxivref{1910.00397}{arxiv:1910.00397}}.

\bibitem{Chaudhuri:1988qb}
S.~Chaudhuri and J.~A.~Schwartz,
\textit{``{A Criterion for Integrably Marginal Operators}''},
\textsf{\doiref{10.1016/0370-2693(89)90393-6}{Phys.~Lett.~B219,~291~(1989)}}.

\bibitem{Klimcik:1995dy}
C.~Klimcik and P.~Severa,
\textit{``{Poisson-Lie T duality and loop groups of Drinfeld doubles}''},
\textsf{\doiref{10.1016/0370-2693(96)00025-1}{Phys.Lett.~B372,~65~(1996)}},
\texttt{\arxivref{hep-th/9512040}{hep-th/9512040}}.

\bibitem{Stern:1998my}
A.~Stern,
\textit{``{Hamiltonian approach to Poisson Lie T - duality}''},
\textsf{\doiref{10.1016/S0370-2693(99)00111-2}{Phys.~Lett.~B450,~141~(1999)}},
\texttt{\arxivref{hep-th/9811256}{hep-th/9811256}}.

\bibitem{Klimcik:2015gba}
C.~Klimcik,
\textit{``{$\eta$ and $\lambda$ deformations as ${\cal E}$-models}''},
\textsf{\doiref{10.1016/j.nuclphysb.2015.09.011}{Nucl.~Phys.~B900,~259~(2015)}},
\texttt{\arxivref{1508.05832}{arxiv:1508.05832}}.

\bibitem{Sfetsos:2013wia}
K.~Sfetsos,
\textit{``{Integrable interpolations: From exact CFTs to non-Abelian
  T-duals}''},
\textsf{\doiref{10.1016/j.nuclphysb.2014.01.004}{Nucl.Phys.~B880,~225~(2014)}},
\texttt{\arxivref{1312.4560}{arxiv:1312.4560}}.

\bibitem{Hollowood:2014qma}
T.~J.~Hollowood, J.~L.~Miramontes and D.~M.~Schmidtt,
\textit{``{An Integrable Deformation of the AdS$_5 \times$S$^5$
  Superstring}''},
\textsf{\doiref{10.1088/1751-8113/47/49/495402}{J.Phys.~A47,~495402~(2014)}},
\texttt{\arxivref{1409.1538}{arxiv:1409.1538}}.

\bibitem{Klimcik:1995ux}
C.~Klimcik and P.~Severa,
\textit{``{Dual nonAbelian duality and the Drinfeld double}''},
\textsf{\doiref{10.1016/0370-2693(95)00451-P}{Phys.Lett.~B351,~455~(1995)}},
\texttt{\arxivref{hep-th/9502122}{hep-th/9502122}}.

\bibitem{Hull:1987yi}
C.~M.~Hull and P.~K.~Townsend,
\textit{``{String Effective Actions From $\sigma$ Model Conformal
  Anomalies}''},
\textsf{\doiref{10.1016/0550-3213(88)90342-2}{Nucl.~Phys.~B301,~197~(1988)}}.

\bibitem{Metsaev:1987zx}
R.~R.~Metsaev and A.~A.~Tseytlin,
\textit{``{Order alpha-prime (Two Loop) Equivalence of the String Equations of
  Motion and the Sigma Model Weyl Invariance Conditions: Dependence on the
  Dilaton and the Antisymmetric Tensor}''},
\textsf{\doiref{10.1016/0550-3213(87)90077-0}{Nucl.~Phys.~B293,~385~(1987)}}.

\bibitem{Meissner:1996sa}
K.~A.~Meissner,
\textit{``{Symmetries of higher order string gravity actions}''},
\textsf{\doiref{10.1016/S0370-2693(96)01556-0}{Phys.~Lett.~B392,~298~(1997)}},
\texttt{\arxivref{hep-th/9610131}{hep-th/9610131}}.

\end{thebibliography}

\end{document}